\documentclass[12pt]{article}
\usepackage{times}
\usepackage{amsmath}
\usepackage{amsfonts}
\usepackage{graphicx,psfrag,epsf}
\usepackage{verbatim}
\usepackage{enumerate}
\usepackage{natbib}
\usepackage{sansmath}
\usepackage{lineno}
\usepackage{tablefootnote}
\usepackage{hyperref}
\usepackage{xcolor}
\usepackage{algorithm}
\usepackage{algorithmic}
\usepackage{tikz}
\usepackage{mwe}
\usetikzlibrary{bayesnet}
\usepackage{forest}
\usepackage{lineno}

\newcommand{\ind}{\stackrel{\rm ind}{\sim}}
\newcommand{\iid}{\stackrel{\rm iid}{\sim}}
\newcommand{\beginsupplement}{
        \setcounter{table}{0}
        \renewcommand{\thetable}{S\arabic{table}}
        \setcounter{figure}{0}
        \renewcommand{\thefigure}{S\arabic{figure}}
}

\newcommand{\alr}{{\rm alr}}
\newcommand{\clr}{{\rm clr}}

\newcommand{\tlr}{{\rm tlr}}

\newcommand{\bmu}{\boldsymbol{\mu}}
\newcommand{\bSig}{\boldsymbol{\Sigma}}

\usepackage[margin=1in]{geometry}
\usepackage{caption}
\usepackage{subcaption}
\usepackage{setspace}

 \usepackage{graphicx}
\setcitestyle{authoryear,open={(},close={)}} 
\begin{document}

  \title{\bf A tree-based model for addressing sparsity and taxa covariance in microbiome compositional count data}
  \author{Zhuoqun Wang \\ Duke University \\ Durham, NC 27708 \and Jialiang Mao \\ LinkedIn Corporation\\ Sunnyvale, CA 94085  \and Li Ma\thanks{Email: \url{li.ma@duke.edu}.} \\ Duke University \\ Durham, NC 27708}
\maketitle

\bigskip

\begin{abstract}
Microbiome compositional data are often high-dimensional, sparse, and exhibit pervasive cross-sample heterogeneity. Generative modeling is a popular approach to analyze such data and effective generative models must accurately characterize these key features of the data. While the high-dimensionality and abundance of zeros have drawn much attention in the literature on modeling microbiome compositions, existing models are often inflexible in their ability to capture the complex cross-sample variability. This would affect statistical efficiency and sometimes lead to misleading conclusions in a variety of inference tasks ranging from differential abundance analysis to the identification of latent structures such as in clustering and network analysis. 
We introduce a generative model, called the ``logistic-tree normal'' (LTN) model, that addresses this need while maintaining the ability to adequately capture other key characteristics of microbiome compositional data such as the abundance of zeros.   
LTN incorporates a tree-based decomposition for effective aggregation over sparse taxa counts and models the relative abundance at the tree splits {\em jointly} using a (multivariate) logistic-normal distribution, which can be endowed with a flexible covariance among taxa as needed in the given application.  The latent Gaussian structure of LTN allows a wide range of multivariate analysis and modeling tools for high-dimensional data---such as those enforcing sparsity or low-rank assumptions on the covariance structure---to be readily incorporated. As a general-purpose, fully generative model and LTN can be applied in a wide range of contexts, while at the same time, efficient computational recipes for Bayesian inference under LTN are available through conjugate blocked Gibbs sampling enabled by P\'olya-Gamma augmentation. 
We demonstrate the use of LTN in a compositional mixed-effects model for differential abundance analysis through both numerical experiments and a reanalysis of the infant cohort in the DIABIMMUNE study.   
We explain and showcase through numerical experiments and the case study how LTN, through adequately accounting for the cross-sample heterogeneity, is capable of generating the appropriate proportion of zeros without incurring an explicit zero-inflation component. This confirms a recent viewpoint that ``zero-inflation'' in count-based sequencing data are often results of unaccounted cross-sample variation.

\end{abstract}

\newpage

\doublespacing

\section{Introduction}
\label{sec:intro}

The human microbiome is the collection of genetic information from all microbes residing on or within the human body. The development of high-throughput sequencing technologies has enabled profiling the microbiome taxa composition in a cost-efficient way through either shotgun metagenomic sequencing or amplicon sequencing on target genes (e.g., the 16S rRNA gene). 
The resulting data are usually a compositional count vector for each sample with the total count of the identified microbes in each sample determined by the sequencing depth of the study and shall only be interpreted proportionally \citep{li2015}. 

The analysis of microbiome studies can involve both supervised and unsupervised tasks. One common task in microbiome studies is to decipher the relationship between microbiome composition and various health outcomes. A common strategy is carrying out a differential abundance analysis by comparing the microbiome compositions of samples collected on two or more groups \citep{nearing2022da,xia2023da}, while adjusting for other contributors to the variability in the taxa abundance, including both measured covariates and additional unmeasured ``random effects''. For example, a core research question in the DIABIMMUNE study \citep{diabimmune}, which is to be reanalyzed in our case study, is to understand the association between infant gut microbiome and Type 1 diabetes (T1D) risk. While a few covariates that are known to contribute to the change in the microbiome compositions are measured, including age and dietary patterns, the compositions display substantial additional variation across the samples beyond what can be explained by these covariates, clearly due to a myriad of other unmeasured causes. Such variability needs to be properly accounted for through statistical formulations---either model-based or model-free---in order to identify the underlying association of interest with valid statistical confidence.  

Another common type of analysis involving microbiome compositions is unsupervised, or exploratory data analysis, aimed at identifying latent structures such as clusters, subcommunities and network structures among microbial taxa. In this context, again one must appropriately differentiate the underlying structure of interest from that due to both measured and unmeasured sources of variability. For example, the key to identifying an underlying cluster of samples lies in distinguishing the within-cluster variability from between-cluster variability. 

A unified approach to the analysis of microbiome compositional data is through building generative models, which allows direct incorporation of key features of such data, including the count nature, the compositionality, the abundance of zeros, as well as the complex cross-sample heterogeneity. A generative model for i.i.d.\ or exchangeable samples of microbiome compositions can serve as building blocks for more sophisticated statistical models to accommodate additional features of the data. This approach has been fruitfully taken by many for both differential abundance analysis \citep{la_rosa2012dm,wadsworth2017,zhang_et_al2020zero,chen2017da,aldex,corncob} and exploratory analysis \citep{holmes2012dm,layeghifard2017network,sankaran2019lda,10242153,fang2023,10.1093/jrsssc/qlac002,10.1093/bioinformatics/btac782,10.1214/21-AOAS1552}. Most of these prior works have given specific attention to the need for accommodating the compositionality and the prevalence of zeros in microbiome data---with a variety of different approaches proposed such as modeling zero-inflation directly \citep{zhang_et_al2020zero,tang2020zero,koslovsky2023zeroDM} as well as using tree-based aggregation to accommodate sparsity \citep{bgcr,bien2021,li2023dart}. With some notable exceptions \citep{mimix,dirfactor}, however, these methods generally adopt restrictive assumptions on the underlying cross-sample variation and often do so implicitly without warning the practitioner who applies them of such constraints. Some recent works have found that the resulting downstream analysis can be sensitive to these implicit assumptions in both differential abundance \citep{dirfactor} and exploratory analyses \citep{shi2022cluster,mao_ma_2022dtmm,leblanc2023}.

 Two popular classes of generative models for microbiome compositions are the log-ratio normal (LN) model \citep{aichison} and the Dirichlet-multinomial (DM) model \citep{holmes2012dm,la_rosa2012dm}.  The LN model can capture rich cross-sample covariance structure among the OTUs; however, when LN is adopted to model the {\em unobserved} relative abundance that gives rise to the observed count vectors, inference is computationally challenging if the number of OTUs is even just moderate ($>50$) due to lack of conjugacy to the multinomial sampling model that is commonly employed to capture the count nature of microbiome compositional data. Consequently, in practice LN is most frequently applied to model the actual {\em observed} proportion of counts (i.e., the counts on the OTUs divided by the total sum of counts for each sample), essentially ignoring the count nature of the sequencing reads.

The DM model, on the other hand, maintains the multinomial sampling model on the counts but instead adopts a Dirichlet distribution on the underlying unobserved relative abundance vector. Dirichlet is conjugate to the multinomial sampling model, and therefore DM is computationally efficient; however, the induced covariance structure under Dirichlet is characterized by a single scalar parameter (which is the sum of the Dirichlet pseudo-count parameters), and thus is way too restrictive for characterizing the cross-sample variability in typical microbiome compositions. A more recent generalization of the DM model, called the Dirichlet-tree multinomial (DTM) model \citep{dennis1991dt,wang&zhao2017} which utilizes an underlying tree structure relating the taxa---usually a phylogenetic tree or a taxonomic tree---lessens this limitation, but only slightly. 
Its cross-sample covariance, though more relaxed than that of the Dirichlet, is still very restrictive as it only uses $(K-1)$ parameters to characterize a $(K-1)\times (K-1)$ covariance structure where $K$ is the total number of taxa. Moreover, the appropriateness of the covariance structure imposed by DTM relies heavily on the extent to which the imposed tree structure (such as the phylogenetic tree) accurately reflects the functional relationship among the taxa in the given context, making the resulting inference particularly sensitive to the specification of the tree, which is a major concern in adopting tree-based methods for microbiome compositions \citep{xiao2018,zhou_bayesian_2021}.

To address these challenges, we propose a new generative model called “logistic-tree normal” (LTN) that combines the key features of DTM and LN models to inherit their respective desired properties.  
LTN utilizes the tree-based decomposition of multinomial sampling model as DTM, which takes advantage of tree-aggregation of sparse counts \citep{wang&zhao2017,washburne2018tree,bien2021,li2023dart}, but unlike  DTM, it employs a latent multivariate LN distribution on the tree splitting probabilities with a general covariance structure. This more flexible covariance structure makes the resulting inference substantially more robust than DTM with respect to the choice of the tree. At the same time, by utilizing a tree-based binomial decomposition of the multinomial model, LTN restores the full conjugacy with the assistance of P\'olya-Gamma (PG) data augmentation \citep{polson}, allowing Gibbs sampling and thereby avoiding the computational difficulty incurred by LN-based models. 

The fully probabilistic, generative nature of LTN allows it to be either used as a standalone model or embedded into more sophisticated models to accomplish both supervised and unsupervised goals.    
The latent multivariate Gaussian formulation allows common modeling and computational techniques for high-dimensional data, such as those based on sparsity and low-rank assumptions on the covariance structure imposed by various types of regularization, to be readily incorporated into LTN while maintaining the computational ease. 

A recent discussion arose in the literature on sequencing count data concerns the source of abundant zeros in count data generated from high-throughput sequencing experiments. A rising viewpoint is that when the data are in the form of sequencing counts, the abundance of zeros can often be adequately captured through appropriately modeling the underlying sources of variation \citep{sarkar_stephens2021,SILVERMANzero}. While these arguments were presented in univariate count models, we believe this viewpoint remains valid in the multivariate context and provide reasoning in the context of LTN on how this model is capable of generating adequate proportions of zeros simply by flexibly modeling the underlying cross-sample variation without an explicit zero-inflation component. This is validated through both numerical experiments in which we examine the ability of LTN to generate zeros and through checking goodness-of-fit of the proportion of zeros in the DIABIMMUNE study.

The rest of the paper is organized as follows. Section 2 introduces LTN after briefly reviewing LN and DTM, examines its distributional properties including its ability to generate sparse data, and presents an LTN-based mixed-effects model tailored for differential abundance analysis. In Section~3, we investigate the performance of the proposed method for differential abundance analysis with several numerical experiments and compare it to two state-of-the-art methods. In Section~4 we carry out a case study on the T1D cohort from the DIABIMMUNE study \citep{diabimmune}. In Section~5 we conclude with some discussion.

\section{Methods}
\subsection{Structure of microbiome compositional count data}
Various bioinformatic preprocessing pipelines such as MetaPhLAN \citep{beghini_integrating_2020} and DADA2 \citep{dada2} have been developed to ``count" the microbes in each sample and report the results in terms of amplicon sequence variant (ASV) or operational taxonomic unit (OTU) abundance.  Both OTU and ASV can serve as the unit for the downstream analysis of microbial compositions and in recent years the ASV has become the standard approach for a number of reasons. However, following the broader tradition in microbial studies, throughout the rest of the paper we shall use the more classic term ``OTU" to refer to either of them.

A typical microbiome dataset consists of an $n\times K$ OTU (count) table with $n$ being the number of samples and $K$ the number of identified OTUs usually after some quality control screening, a taxonomic table summarizing the taxonomic identification of the identified OTUs to known or unknown taxa at multiple taxonomic levels, as well as possibly a set of covariates on the samples measured by medical monitoring, questionnaires or other means. One can usually also construct a summary of the evolutionary relationship of OTUs in the study in the form of a phylogenetic or taxonomic tree. See \cite{li2015} for additional details and examples.  

Some common characteristics of microbiome compositional data that need to be accounted for in the modeling and analysis of such data include (i) high-dimensionality---large $K$, (ii) sparsity---many counts in the OTU table are zeros, and (iii) pervasive cross-sample heterogeneity---the OTU count vectors usually display large variability across the samples and involve complex covariance among the OTUs. We note that the sparsity in microbiome data can come in two forms: (1) with-sample sparsity---each microbiome count sample have non-zero counts in a small proportion of identified OTUs, and (2) within-OTU sparsity---most OTUs are observed in only a small number of samples. These two types of sparsity are usually both present in microbiome datasets. While the key motivation for our proposed method is in the high-dimensionality and the complex cross-sample covariance, we will provide examples in our later numerical experiments and case study to show how our proposed model effectively accounts for both types of sparsity.

\subsection{Two popular generative models for microbiome compositional data}
\begin{figure}
     \centering
     \begin{subfigure}[b]{0.4\textwidth}
         \centering
         \includegraphics[width=\textwidth]{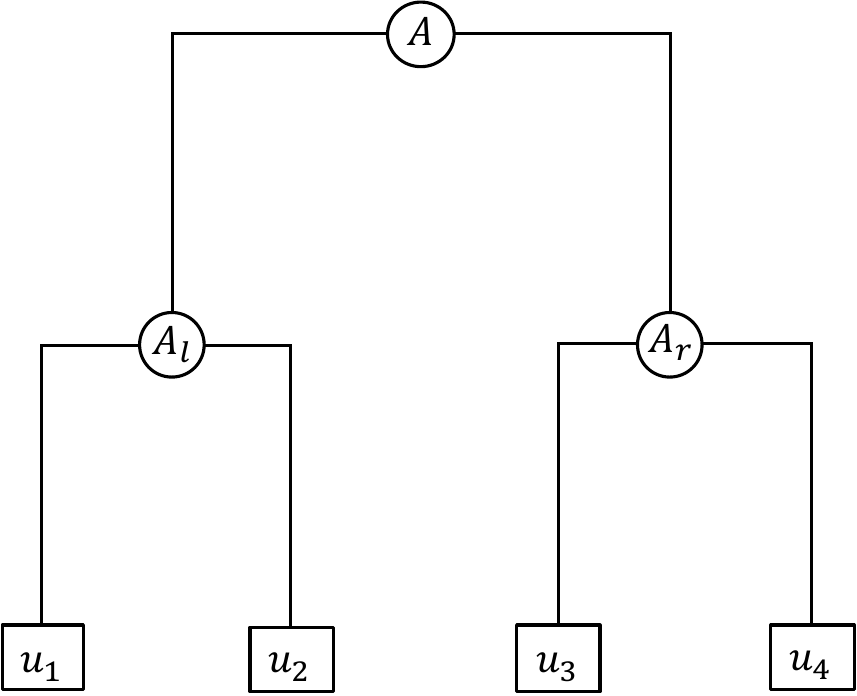}
         \caption{\textit{Notation for nodes}}
         \label{fig:phy_tree}
     \end{subfigure}
     \hfill
     \begin{subfigure}[b]{0.5\textwidth}
         \centering
         \includegraphics[width=\textwidth]{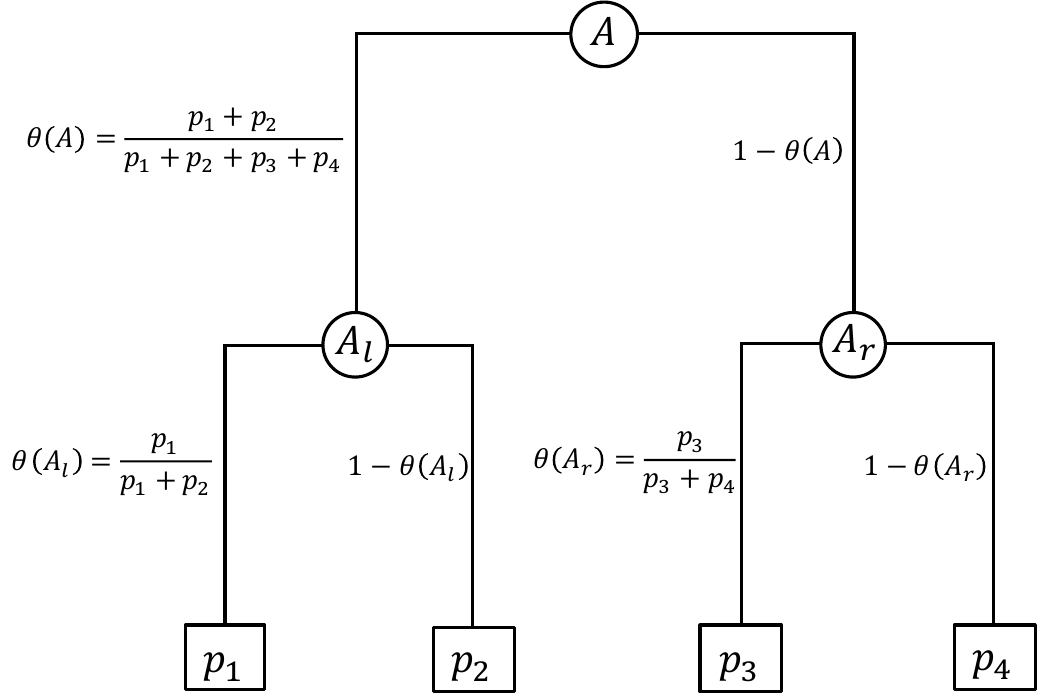}
         \caption{\textit{Tree-based transform from $\boldsymbol p$ to $\boldsymbol \theta$}}
         \label{fig:tlr}
     \end{subfigure}
        \caption{An example of a phylogenetic tree over four OTUs}
\end{figure}

We will first briefly review two most widely used classes of generative models for OTU counts, namely log-ratio normal (LN) and Dirichlet-multinomial (DM) models. They inspire our new model to be introduced later. We start by introducing some basic notation that will be used throughout the paper. Suppose there are $n$ samples and a total of $K$ OTUs, denoted by $u_1,u_2,\ldots,u_K$. The OTU table is an $n\times K$ matrix, whose $i$th row represents the OTU counts in sample $i$ and the $(i,j)$-th element represents the count of OTU $j$ in that sample.  Let $\mathcal{T}=\mathcal{T(I,U;E)}$ be a rooted full binary phylogenetic tree over the $K$ OTUs, where $\mathcal{I,U,E}$ denote the set of interior nodes, leaves, and edges respectively. Formally, we can represent each node $A$ in the phylogenetic tree by the set of its descendant OTUs. Specifically, for a leaf node $A\in\mathcal{U}$, which by definition contains only a single OTU $u$, we let $A=\{u\}$. Then the interior nodes can be defined iteratively from leaf to root. That is, for $A\in \mathcal{I}$ with two children nodes $A_l$ and $A_r$, we simply have $A=A_l\bigcup A_r$. 
Figure~\ref{fig:phy_tree} shows an example of a node $A$ containing four OTUs, $\{u_1,u_2,u_3,u_4\}$. 

Suppose $\boldsymbol X=(X_1,\cdots,X_K)'$ is the OTU counts for a sample, and  $N=\sum_{j=1}^K X_{j}$ is the total number of OTU counts. A natural sampling model for the OTU counts given the total count $N$ is the multinomial model 
 \begin{align}
 \label{eq:multinom}
 \boldsymbol{X}\,|\,N,\boldsymbol{p}\sim{\rm Multinomial}(N,\boldsymbol{p}),
 \end{align}
 where $\boldsymbol{p}=(p_{1},\cdots,p_{K})'$ is the underlying OTU (relative) abundance vector, which lies in a $(K-1)$-simplex. That is, $\boldsymbol{p}\in \mathcal{S}_{K}=\{(p_1,\cdots,p_K)':p_j\geq 0,j=1,\cdots, K, \sum_{j=1}^K p_j=1\}$. The simplicial constraint on relative abundances can present modeling inconveniences, and a traditional strategy is to apply a so-called log-ratio transformation to map the relative abundance vector into a Euclidean space \citep{aichison}.
 
Three most popular choices of the log-ratio transform are the additive log-ratio (alr), centered log-ratio (clr) \citep{aichison} and isometric log-ratio (ilr) \citep{egozcue_isometric_2003,egozcue_changing_2016}, all of which have been applied to microbiome compositional data in the literature. For a composition $\boldsymbol{p}$, the clr and alr transforms are given by
\begin{align*} 
\clr(\boldsymbol{p})=\{\log(p_j/g(\boldsymbol{p})):j=1,\cdots,K\}\quad \text{and} \quad \alr(\boldsymbol{p})=\{\log(p_j/p_K):j=1,\cdots,d\}
\end{align*}
where $d=K-1$ and $g(\boldsymbol{p})$ is the geometric mean of $\boldsymbol{p}$. The ilr transform, on the other hand, uses a binary partition tree structure to define the log-ratios, and it turns $\boldsymbol p$ into ``balances" associated with the interior nodes of the tree. The balance $\eta(A)$ associated with an interior node $A$ of the tree is defined as
$$\eta(A)=\sqrt{\frac{|A_l||A_r|}{|A_l|+|A_r|}}\log \frac{g(\boldsymbol{p}(A_l))}{g(\boldsymbol{p}(A_r))},$$
where $|A_l|$ and $|A_r|$ are the number of OTUs in the left and right subtree of node $A$ respectively, and $g(\boldsymbol{p}(A_l))$ and $g(\boldsymbol{p}(A_r))$) are the corresponding geometric means of the probability compositions of leaves in the left and right subtrees of node $A$. While in many applications finding a suitable binary tree for the ilr might not be easy, in the microbiome context, a phylogenetic tree or a taxonomic is a common choice \citep{philr}.

{\em Log-ratio normal (LN)} models posit that the log-ratios computed from these transforms are multivariate Gaussian. 
These models have been successfully used in characterizing microbial dynamics \citep{aijo,mallard} and linking covariates with microbiome compositions \citep{xia_logistic_2013,mimix}. Inference using the LN models can incur prohibitive computational challenges when the number of OTUs grow due to the lack of conjugacy between the multinomial likelihood and multivariate normal, and recent efforts have been made to overcome this computational challenge through approximate computation. Interested readers can find a detailed discussion on the computational challenges incurred by LN and possible approaches to cope with them in \cite{silverman2022}.

{\em Dirichlet} models are another classical model for the compositional vector $\boldsymbol{p}$ that is conjugate to the multinomial sampling model. For this reason they are often referred to as {\em Dirichlet-multinomial (DM)} models in the microbiome modeling literature \citep{la_rosa2012dm,holmes2012dm}. 
DM models are computationally more efficient than LN, and are also capable of generating very sparse count vectors.
However, this model class affords only a single scalar parameter, which is the sum of the Dirichlet pseudo-count parameters, to characterize the covariance structure among the OTUs, essentially assuming that all OTUs are mutually independent modulo the artificial dependence caused by the compositional constraint $\sum_j p_{j}=1$. This is clearly too restrictive for typical microbiome datasets where taxa exhibit complex covariance structures \citep{layeghifard2017network,coat,matchado2021network,faust2021network}. 

{\em Dirichlet-tree multinomial (DTM)} models \citep{dennis1991dt} have been introduced to alleviate this limitation of the Dirichlet while maintaining its computational tractability. Given a dyadic tree $\mathcal{T}$ over the OTUs, e.g., a phylogenetic or taxonomic tree, DTM utilizes the fact that multinomial sampling is equivalent to sequential binomial sampling down a binary partition. Specifically, the multinomial sampling model in Eq.~\eqref{eq:multinom} is equivalent (in the sense that the likelihood function is identical up to a normalizing constant) to independent binomial sampling models defined on the interior nodes of $\mathcal{T}$. That is,
$$
y(A_l)\,|\,y(A),\theta(A) \ind{\rm Binomial}(y(A), \theta(A))\qquad\text{for all }A\in\mathcal{I},
$$
where $y(A)=\sum_{j:u_j \in A} X_{j}$ is the total counts from OTUs in the subtree rooted at $A$ and $\theta(A) =\frac{\sum_{j:u_j\in A_l}p_j}{\sum_{j:u_j\in A}p_j}$ is relative abundance of the left subtree of $A$ with respective to $A$. See Figure~\ref{fig:tlr} for an illustration.

Under this perspective, the relative abundance vector is now represented using the collection of binomial ``branching'' probabilities $\theta(A)$ for all $A\in\mathcal{I}$, the interior nodes of the tree. DTM adopts a beta model for each $\theta(A)$ to attain conjugacy with respect to the binomial likelihood. That is,
\[
\theta(A) \ind {\rm Beta}(\mu(A)\nu(A),(1-\mu(A))\nu(A))\qquad\text{for }A\in\mathcal{I},
\]
where $\mu(A)$ specifies the mean of the branching probabilities, and $\nu(A)$ is the concentration parameter that controls the variance for any given mean.

Tree-based aggregation has been employed in several different approaches to analyze microbiome data, usually by using the phylogenetic or taxonomic trees due to their ready availability. Some examples include the ilr-based LN approach mentioned above, as well as a number of other non-generative tree-aggregation approaches  \citep{washburne2018tree,bien2021,li2023dart}. A distinct feature in DTM differentiates it from all of the other tree-based methods including the ilr-based LN model. 
It is the fact that DTM not only employs a tree-based transform of the compositional abundance vector $\boldsymbol p$ as the ilr does, 
but it also decomposes the {\em sampling model} (i.e., the multinomial likelihood) into bionomial models, which is key to achieving its computational tractability through the beta-binomial conjugacy.

Nevertheless, DTM does not satisfactorily resolve the limitation of DM in terms of cross-sample variation. To see this, note that the parameters $\nu(A)$'s are the only ones that characterize the covariance across the $K$ OTUs. In fact, DTM still assumes mutual independence among $\theta(A)$'s and so there are a total of only $K-1$ scalar parameters, which along with the underlying phylogenetic tree characterize the covariance among the $K$ OTUs. This also makes inference with DTM particularly sensitive to the phylogenetic tree. In contrast, LN models, including the tree-based ilr models, allow flexible specification of the $(K-1)\times (K-1)$ covariance structure. This allows ilr to be less sensitive to misspecification of the dyadic tree. 

Moreover, even though DTM is fully generative, it is actually not easy to embed DTM into more sophisticated hierarchical models. This is because there is no known (hyper)prior for beta distributions, and so DTM will incur computational challenges should there be additional modeling components, such as covariate effects on its mean and priors on the unknown concentration parameters. Those features will also break the conjugacy, resulting in the need for numerical integration \citep{christensen_ma_2020,mao_ma_2022dtmm}.

\subsection{Logistic-tree normal models}

We introduce a generative model that combines the tree-based factorization of the multinomial sampling model under DTM {\em with} the log-ratio transform under LN, thereby enjoying key benefits of both models while addressing all of their aforementioned limitations. In particular, our model achieves both flexibility in characterizing taxa covariance \textit{and} computational tractability, while maintaining the ability to accommodate sparsity. Specifically, we adopt a log-ratio transform (in this case simply a logistic transform) on the binomial branching probability $\theta(A)$ on each interior node $A$ \citep{jara&hanson:2010}. That is, we model the log-odds on the interior nodes 
\[
\psi(A)=\log\frac{\theta(A)}{1-\theta(A)} \qquad \text{for all }A\in \mathcal{I}
\]
jointly as a multivariate Gaussian.

More formally, for a compositional probability vector $\boldsymbol{p}=(p_1,p_2,\ldots,p_K)\in \mathcal{S}_{K}$, we define the {\em tree-based log-ratio} (tlr) transform as $\tlr(\boldsymbol{p})=\boldsymbol\psi=\{\psi(A):A\in\mathcal{I}\}$, where 
$\psi(A)=\log\frac{\theta(A)}{1-\theta(A)}$. 
Like other log-ratio transforms, tlr maps a compositional vector $\boldsymbol{p}$ to a vector of $d$ log-odds $\psi(A)$. 
Finally, modeling the tlr$(\boldsymbol{p})$ as MVN$(\boldsymbol{\mu},\boldsymbol{\Sigma})$, we have the full formulation of a {\em logistic-tree normal} (LTN) model: 
\begin{equation}\label{eq:generate}
    \begin{split}
y(A_l)|y(A),\theta(A) & \ind {\rm Binomial}(y(A),\theta(A))\qquad\text{for } A \in \mathcal{I},\\\
\psi(A)&=\log \frac{\theta(A)}{1-\theta(A)}\qquad \text{for } A \in \mathcal{I},\\
\boldsymbol{\psi} & \sim {\rm MVN}(\boldsymbol{\mu},\boldsymbol{\Sigma}).
\end{split}
\end{equation}
 We denote this model by LTN$(\boldsymbol\mu,\boldsymbol\Sigma)$. A graphical model representation of the LTN model for a dataset with $n$ exchangeable samples indexed by $i = 1,\cdots,n$ is shown in Figure \ref{fig:original}. 

At first glance, while LTN affords flexible mean and covariance structures, it suffers from the same lack of conjugacy as existing LN models based on log-ratio transforms such as alr, clr and ilr. Fortunately, the binomial decomposition of the likelihood under LTN allows a data-augmentation technique called P\'olya-Gamma (PG) augmentation \citep{polson} to restore the conjugacy. In fact, this data-augmentation strategy ensures full conjugacy for LTN models even with additional modeling hierarchy such as covariates and multiple variance components, provided that one adopts any conjugate (hyper)priors on the Gaussian parameters $(\boldsymbol{\mu},\boldsymbol{\Sigma})$, making LTN substantially more versatile and computationally tractable than DTM. A graphical model representation of LTN with PG augmentation for a dataset of $n$ exchangeable samples is shown in Figure~\ref{fig:ltn_pg}. Details of the P\'olya-Gamma augmentation for LTN are provided in Supplementary Material \ref{sec:s-pg}. 

\begin{figure}
     \centering
     \begin{subfigure}[b]{0.45\textwidth}
         \centering

         \tikz{ 
                    
    \node[latent] (mu) {$\boldsymbol{\mu}$} ; 
    \node[latent,above=of mu] (omega) {$\boldsymbol{\Sigma}$} ; 
    \node[latent,right=of mu] (psi) {${\psi}_i(A)$} ;  
   \node[obs, right=of psi] (yl) {$y_i(A_l)$} ; 
   \node[latent, above=of yl] (y) {$y_i(A)$} ;
   \plate[inner sep=0.2cm, xshift=-0.1cm, yshift=0.12cm] {plate1} {(yl)(y)(psi)}
   {$A\in\mathcal{I}$}; 
     \plate[inner sep=0.2cm,xshift=0.1cm,yshift=0.1cm] {plate2} {(plate1)}
   {$i=1,\cdots,n$}; 
   \edge {mu} {psi} ; 
   \edge {omega} {psi} ; 
    \edge {psi} {yl} ; 
    \edge {y} {yl} ; 
  }

  \caption{original model}
         \label{fig:original}
     \end{subfigure}
     \hfill
     \begin{subfigure}[b]{0.45\textwidth}
         \centering
         \tikz{ 
    \node[latent] (mu) {$\boldsymbol{\mu}$} ; 
    \node[latent,above=of mu] (omega) {$\boldsymbol{\Sigma}$} ; 
    \node[latent,right=of mu] (psi) {${\psi}_i(A)$} ;  
   \node[obs, right=of psi] (yl) {$y_i(A_l)$} ; 
   \node[latent, above=of yl] (y) {$y_i(A)$} ;
     \node[latent, above=of psi] (pg) {$w_i(A)$} ; 
   \plate[inner sep=0.2cm, xshift=-0.1cm, yshift=0.12cm] {plate1} {(yl)(y)(psi)}
   {$A\in\mathcal{I}$}; 
     \plate[inner sep=0.2cm,xshift=0.1cm,yshift=0.1cm] {plate2} {(plate1)}
   {$i=1,\cdots,n$}; 
   \edge {mu} {psi} ; 
   \edge {omega} {psi} ; 
    \edge {psi} {yl} ; 
    \edge {y} {yl} ; 
    \edge{y} {pg} ; 
    \edge{psi} {pg} ; 
  }
  \caption{with P\'olya-Gamma augmentation: \\$w_i(A)|y_i(A),\psi_i(A)\ind {\rm PG}
  (y_i(A),\psi_i(A))$ is introduced to restore conjugacy}
         \label{fig:ltn_pg}
     \end{subfigure}
        \caption{A graphical model representation of LTN for $n$ exchangeable samples}
        \label{fig:ltn}
\end{figure}

\subsection{Distributional properties of LTN and the generation of zeros}

Before we consider applications of LTN to inference tasks, we first examine two important distributional properties of LTN relevant to its ability to accommodate common features of microbiome compositional data. In particular, while the design of LTN is aimed at enriching the taxa covariance and maintaining scalability with respect to dimensionality, two important questions remain---(i) how capable are LTN in generating {\em sparse} data---displaying both within-OTU {\em and} within-sample sparsity; and (ii) how sensitive is the resulting analysis with respect to the choice of the dyadic tree? We provide some reasoning here and will revisit these considerations later in our numerical experiments and the case study to corroborate our argument with empirical evidence.

{\em Characterizing abundance of zeros.} A key feature of microbiome compositional data is the vast numbers of zero counts typically observed when the data are presented at deep taxonomic levels. The insufficiency of standard generative models for count-valued data to accommodate such pervasive sparsity has been noted in the literature and many recent developments focus on incorporating a zero-inflation component to accommodate this aspect of the data \citep{zhang_et_al2020zero,tang2020zero,koslovsky2023zeroDM}. We acknowledge the importance of accommodating zeros in the data, but share with some others \citep{sarkar_stephens2021,silverman2020_zero} the concern that an artificial zero-inflated component in the generative model, except under the scenario that it is justified by either the specific biological mechanism or experimental protocol, can distort the identifiability and interpretability of the actual meaning of the ``non-zero'' component in such models, leading to bias in the downstream inference. Hence we would, except in some special cases, refrain from introducing an explicit zero-inflation component into LTN-based models, even though they are not technically difficult to construct.

Next we argue that the design of LTN allows the generation of highly sparse counts without an explicit zero-inflation component, and the reasons are different for within-OTU and with-sample sparsity. 
First, we note that within-OTU sparsity---that is, a large proportion of samples have zero counts for an OTU---arises when the actual relative abundance of that OTU is low over the majority of samples. This scenario can be readily captured under LTN when a sub-branch of the tree that contains the OTU under consideration has a small branching probability. Specifically, within-OTU sparsity can therefore arise when the mean parameter $\mu(A)$ is large (in absolute value) on any ancestral node $A$ of the OTU of interest. On the other hand, within-sample sparsity---that is, a sample of count vector has many zeros and the OTUs with zero counts vary wildly across samples---arises for a different reason. They arise due to very large cross-sample variability which can be readily induced by large marginal variance of the log-odds, $\psi(A)$, on the interior nodes $A$. It is important to note that LTN can therefore accommodate both the within-OTU and with-sample sparsity {\em without} changing the underlying correlation structure among the taxa, which is critical in ensuring the identifiability of the resulting taxa correlation. 

\begin{figure}[h]
    \centering
    \includegraphics[width=0.9\linewidth]{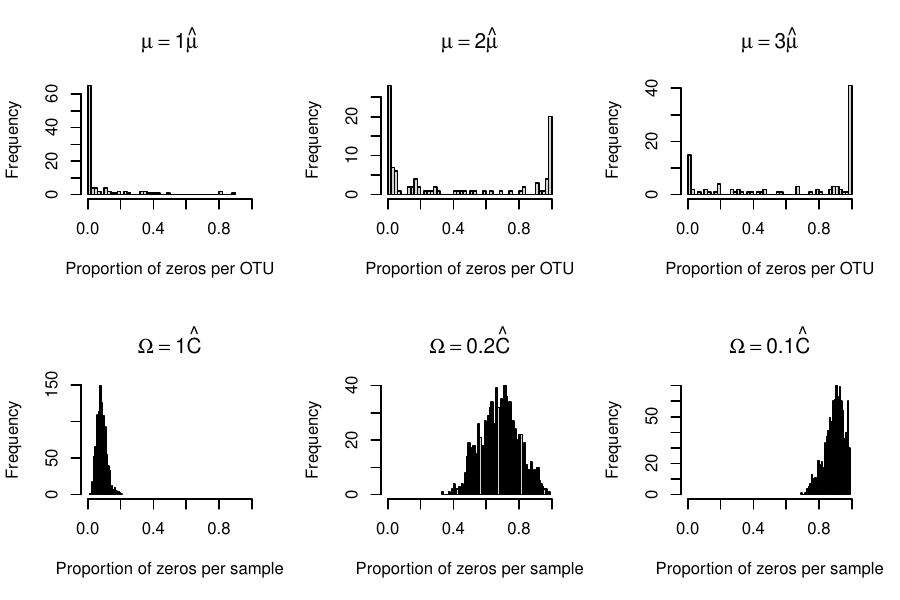}
    \caption{Proportion of zeros by OTU (top) and by sample (bottom). Top: Samples are generated from LTN$(\mu,\hat{C})$ with fixed mean and correlation and varying marginal variance. Bottom: Samples are generated from LTN$(\hat{\mu}, \Omega^{-1})$ with fixed covariance and varying mean. The baseline parameter values $\hat{\mu}$ and $\hat{C}$ are the estimated $\mu$ and partial correlation matrix from an individual in the DIABIMMUNE data. The phenomenon holds generally and is not unique to the particular individual.}
    \label{fig:mu_omega_scaling}
\end{figure}

While more extensive numerical evidence will be provided later in our simulation and case study sections, here we provide a simple numerical example to demonstrate the effect of changing the mean and variance of the log-odds on the interior nodes under LTN and their impact on the within-OTU and within-sample sparsity to support the argument above. We generate samples from a simple LTN model whose parameters $\hat{\bmu}$ and $\hat{\bSig}=\hat{C}^{-1}$ are estimates from the DIABIUMME data (to be discribed later) \citep{diabimmune} under a simplified model that assumes all samples are exchangeable. We then modify the values of $\bmu$ and $\bSig$ by applying different scaling to $\hat{\bmu}$ and $\hat{\bSig}$, thereby adjusting the absolute values of $\bmu$ and the marginal variance of the $\psi(A)$'s which respectively leads to different levels of within-OTU and within-sample sparsity as presented in Figure~\ref{fig:mu_omega_scaling}. This simulation is simplistic and only serves the purpose of demonstrating the general ability for LTN to generate within-OTU and within-sample sparsity. We will examine the goodness-of-fit of the LTN model to the zero patterns observed in the DIABIMMUNE study at a per-sample level when we construct a more sophisticated LTN-based mixed-effects model tailored for the actual study design of the DIABIMMUNE project in our case study. (See Figure~\ref{fig:zeros_diabimmune}.)

{\em Robustness to the choice of the tree.} The phylogenetic and taxonomic trees are generally available for microbiome data as they are constructed from the genetic sequencing reads, but using them out of convenience in the modeling or analysis of  microbiome compositions raises some natural concerns. After all, algorithmically reconstructed evolutionary trees may not necessarily be good proxies to the functional relationship of the OTUs in a given context. This can happen in two ways. First, the evolutionary relationships of the taxa, even if reconstructed perfectly, may not correspond well to their functional relationships. For example, due to selective pressure, OTUs that are far from each other evolutionarily may actually be functionally similar or closely related. Ideally, a functional tree tailored for a given phenotype of interest should be constructed and used for modeling and analysis, but this is usually difficult in most applications. Second, even in cases when the evolutionary relationship among the OTUs does approximate their functional relationship closely, there is often substantial uncertainty in the reconstruction of the underlying evolutionary tree based on sequencing reads, and different bioinformatic algorithms for tree construction can often result in trees that are different in non-trivial ways. 

Therefore an important question for all tree-based approaches to analyzing microbiome data is---what if the chosen tree provides a poor characterization of the functional relationship of the taxa in the given dataset? In constructing and applying such methods, the sensitivity with respect to the choice of the tree should be examined. Here we first provide some conceptual reasoning of the improved robustness of LTN, compared to other tree-based methods such as DTM, with respect to the tree, after which we will provide  numerical evidence to support the reasoning. 

The choice of the tree structure could potentially affect the downstream inference under a tree-based model through inflated uncertainty and/or inflated bias. First, note that the choice of the tree leads to different parametrization of the multinomial mean $\boldsymbol{p}$. An effective re-parametrization of $\boldsymbol{p}$ into the log-odds $\psi(A)$ should ideally involve a small number of $\psi(A)$ with large ``structural variability''. (Here ``structural variability'' corresponds to the variation of interest, such as the cross-group difference in differential abundance analysis and the cross-cluster difference in clustering.)  This is in essence the ``energy concentration'' phenomenon observed in other tree-based transforms such as Haar wavelets. To understand this in the context of microbiome studies, consider the following idealistic scenario in which two taxa (which can be any taxonomic level) are functionally identical and therefore vary together with perfect correlation. In that case, a tree that places them as siblings of a shared parent node will have the same splitting probabilities at the parent node across {\em all} samples. In particular, if one is interested in carrying out a differential abundance analysis across two conditions then there would be no differential splitting at that parent tree node. At the same time, the signal-to-noise ratio for differential splitting at the grandparent node of two taxa will be large due to the tree-based aggregation that combines the counts of the two relevant taxa into a single branch, achieving ``energy concentration'' in very much the same way as Haar wavelet transforms. The tree that most effectively characterizes the functional similarity would then perfectly split such functionally identical taxa into siblings without first breaking apart either into smaller sub-taxa, which leads to ``energy dilution''. In reality, the adopted phylogenetic or taxonomic tree will never be such a perfect functional tree and there are also no such perfectly correlated taxa. Poor choices of the tree, on the other hand, would do the opposite, causing pervasive ``energy dilution'' by dispersing the counts of OTUs that are functionally similar over many different splits of the tree and therefore causing a loss of statistical sensitivity.
In the context of Bayesian inference, this generally demonstrates itself in terms of larger posterior uncertainty, whereas from a sampling perspective, increased variance in the decision rules.

Now let us turn to the second issue of inflated bias, which we argue that LTN is generally robust against while some existing tree-based models such as DTM suffer significantly. Misspecified trees can introduce substantial bias into the resulting inference if it injects strong and unreasonable constraints on the underlying mean and covariance of the taxa distribution. This is true in the case of DTM as it assumes that the tree splitting probabilities are essentially mutually independent, aside from the compositionality constraint. In other words, such mutual independence will only occur in practice when the phylogenetic tree essentially is a {\em perfect} tree that characterizes the functional relationship so that the correlation between any two taxa is exactly described by their distance in the tree, which in almost all cases is biologically implausible. 

In contrast, LTN does not constrain the underlying covariance structure based on the tree (though such constraint could be incorporated through priors on the covariance). Therefore LTN supports all possible covariance even when the tree structure is misspecified. A poor choice of the tree in LTN would generally still lead to a loss of efficiency (increased posterior uncertainty and standard errors) but generally not large bias (unless of course the prior specification on LTN imposes strong constraints based on the tree).

We carry out a somewhat simplistic simulation experiment to demonstrate these points and report the details in Supplementary Material \ref{sec:s-treesensitivity}. Specifically, we generate LTN data from a given tree, but estimate the induced mean and covariance of the OTUs using severely misspecified tree. We then examine the resulting inference on the mean $\bmu$ and covariance $\bSig$ of the log-odds under correctly and misspecified trees. Our sensitivity analyses show that the inference of the mean and covariance structure based on LTN with misspecified trees is generally robust even when the tree is grossly misspecified. 

We will provide additional evidence for the robustness of LTN with respect to tree misspecification under the context of differential abundance analysis in our simulations studies reported in Section~\ref{sec:numerical_examples}, in which we generate synthetic associations between the taxa and an outcome of interest without regard to the underlying tree structure adopted by LTN, and yet LTN leads to competitive performance in identifying the differential abundance.

\section{Application in differential abundance analysis}
\subsection{A mixed-effects regression model for microbiome compositions}
 \label{sec:mixed_effects}
 
So far, we have focused on the sampling properties of LTN as a general-purpose generative model for microbiome compositional counts. Next we turn to its application in practice. While in principle LTN can serve as a prior for a single compositional abundance vector, we believe that in most applications, such as a typical microbiome study, one would employ LTN as the sampling model for multiple samples of abundance vectors and so this will be the focus of our following exposition. Aside from the computational considerations, inference with LTN can in principle proceed from either a Bayesian or a non-Bayesian strategy (through for example maximum likelihood estimation) on learning the parameters $\bmu$ and $\bSig$. Here we shall focus on the Bayesian approach by constructing hierarchical models that adopt LTN as a sampling module for (conditionally) exchangeable microbiome compositional samples. 

There are myriad ways in which one might construct such hierarchical models. We illustrate through  constructing a mixed-effects model given the broad applicability of such models in microbiome studies and focus on the common task of detecting differential abundance of microbial taxa across two conditions (e.g., cases versus controls etc.). Mixed-effects models \citep{mimix,dirfactor} can readily incorporate common design features such as covariates, batch effects, multiple time points, and replicates at different levels. We consider a common study design where a dataset involves $n$ samples that can be categorized into two contrasting groups (e.g., case vs control, treatment vs placebo, etc.), and the practitioner is interested in testing whether there is cross-group difference in microbiome compositions. For each microbiome composition sample $i$, let $s_i\in \{0,1\}$ be the group indicator and $\boldsymbol z_i$ a $q$-vector of covariates.

We consider random effects that involve measured subgrouping structure (e.g., individuals). Specifically, we assume that there is a random effect for each sample $i$ associated with an observed grouping $g_i\in \{1,\cdots,G\}$. For example, in longitudinal studies, $g_i$ can represent the individual from which the $i$th sample is collected. We associate the covariates and random effects with the microbiome composition through the following linear model:
\begin{equation}
    \boldsymbol\psi_i=\boldsymbol\alpha s_i+\boldsymbol\beta^T \boldsymbol z_i+\boldsymbol\gamma_{g_i}+\boldsymbol\epsilon_i,
    \label{eq:mixed_effects}
\end{equation}
where $\boldsymbol\alpha$ represents the potential cross-group differences, $\boldsymbol\beta$ is a $q\times d$ matrix of unknown fixed effect coefficients of the covariates, $\boldsymbol\gamma_{g_i}$ is the random effect from individual $g_i$, which we model exchangeably as $\boldsymbol \gamma_g\iid {\rm MVN}(\boldsymbol0,\boldsymbol\Sigma)$ for $g=1,\cdots,G$. 

Regarding the remaining ``noise'' $\boldsymbol\epsilon_i$, it characterizes the additional variability in the microbiome composition beyond the measured covariates. A simple choice is to assume independent noise $\epsilon_i(A)\ind {\rm N}(0,\sigma_\epsilon^2(A))$ over the tree nodes $A$, and one can adopt inverse-Gamma prior on $\sigma_\epsilon^2(A)$. One may question the independent noise assumption as additional perturbations in the microbiome usually also involve OTU correlation. In adopting independent noise over the tree nodes, we are making the simplifying assumption that all remaining OTU covariance is captured under the phylogenetic tree structure. One can further relax this assumption by incorporating a more flexible OTU covariance in $\epsilon$. One strategy is to adopt a latent factor model that assumes makes a low-rank assumption on the noise covariance. This can in fact also be readily accommodated by modifying the model in \eqref{eq:mixed_effects} to the following
\begin{equation}
    \boldsymbol\psi_i=\boldsymbol\alpha s_i+\boldsymbol\beta^T \boldsymbol z_i+\boldsymbol\gamma_{g_i}+\boldsymbol\lambda^T \boldsymbol x_i+\boldsymbol\epsilon_i,
    \label{eq:mixed_effects_latent_factor}
\end{equation}
where $\boldsymbol x_i$ represents the latent factors for sample $i$ with $\boldsymbol{\lambda}$ being the corresponding loadings. Additional modeling could be employed to accommmodate the spatial-temporal patterns in the observations by modeling the latent factors using a spatial-temporal process. For expositional simplicity we do not further discuss the details of this latent factor model.

A remaining part of the model needs to be specified is the prior on the covariance matrix $\bSig$. When the number of OTUs $K$ is large, proper regularization contraints must be enforced to ensure the model is still identifiable and can be reliably inferred.   
Existing approaches generally enforce either low-rank assumptions or sparsity on the covariance or its inverse, the precision matrix. A popular prior on the precision matrix $\boldsymbol\Omega =\bSig^{-1}$ is to enforce some shrinkage toward a sparse structure. Well-known examples of such models on Gaussian graph include the graphical-Lasso (gLasso) \citep{wang2012} and the graphical horseshoe (gHS) \citep{li2019ghs} priors. Both of these models can be readily employed under LTN. For illustration, in our following numerical examples, we adopt the gLasso prior on $\Omega$, which takes the form
$$    p(\boldsymbol\Omega|\lambda)=C^{-1}\prod_{j<j'}\{{\rm DE}(\omega_{jj'}|\lambda)\}\prod_{j=1}^d\{{\rm EXP}(\omega_{jj}|\lambda/2)\} {1}_{\boldsymbol\Omega \in M^{+}}, $$
where DE$(x|\lambda)=\frac{\lambda}{2} {\exp}(-\lambda|x|)$ is the double exponential density function which essentially imposes an $L_1$ regularization, EXP$(x|\lambda)=\lambda\exp(-\lambda x)1_{x>0}$ is the exponential density function, $M^+$ is the space of positive definite matrices, and $C$ is the normalizing constant that does not involve $\boldsymbol\Omega$ or $\lambda$. Posterior sampling on this model can be conducted using a blocked Gibbs sampling scheme proposed in \cite{wang2012}. With the P\'olya-Gamma augmentation, all parts of the model can be drawn from conjugate full conditionals. The details of the sampler are provided in Supplementary Materials \ref{sec:s-sampler}. 

We note that the sparse or shrinkage priors such as gLasso and gHS are but one of many choices that one can adopt. A natural alternative is again latent factor models that impose low-rank constraints on the covariance matrix, which can also be readily incorporated here without complicating the inference algorithms. We choose the gLasso prior for two reasons. First, to our knowledge priors that impose sparsity constraints have not been previously exploited in the context of microbiome modeling. Second and more importantly, given the analogy of the tree-based decomposition into binomial experiments to that of Haar wavelet transforms, one may naturally expect a similar ``whitening effects'' of the decomposition as commonly observed in wavelet transforms on correlated Gaussian observations by which the (conditional) dependence of the ``wavelet coefficients'', here the log-odds $\psi(A)$, are weakened resulting in a sparse precision matrix. 

Differential abundance analysis usually involves hypothesis testing on the two-group difference and identifying the differentially abundant taxa. Under the above model, this can be accomplished through testing a collection of hypotheses on the interior nodes:
$$H_0: \alpha(A)=0 \text{ versus } H_1: \alpha(A)\neq 0\qquad \text{for }A \in \mathcal{I}$$
To perform these local tests, we adopt a Bayesian variable selection strategy and place the following spike-and-slab priors on entries of $\boldsymbol\alpha$:
\begin{align*}
    \alpha(A)|\pi(A),\phi_\alpha&\ind (1-\pi(A))\delta_0+\pi(A)(1-\delta_0){\rm N}(0,1/\phi_\alpha)\qquad\text{for }A\in \mathcal{I},\\
    \pi(A)&\iid {\rm Beta}(m,1-m)\qquad\text{for }A\in \mathcal{I},\\
    \phi_\alpha&\sim {\rm Gamma}(t,u),
\end{align*}
where $\delta_0$ is point mass at 0, and $m,t,u$ are pre-specified hyperparameters. In particular, $m$ controls the prior marginal probability for the alternative on node $A$, and therefore it plays a critical role in controlling for multiple testing. There are generally two approaches in choosing $m$, either by directly setting $m$, which corresponds to the prior expected proportion of nodes on which the alternative is true, or by setting $m$ so that the induced prior probability for the global null (and alternative) at a desired level. The first approach is less stringent and is more suitable when one desires to identify taxa that are differentially abundant while controlling false discovery rate (FDR). The second is more suited when the goal is to test for the global null.
We adopt continuous priors on the coefficients of the covariates: $\boldsymbol{\beta}\sim {\rm N}_{q\times d}(0, cn \boldsymbol{I}_d \otimes (\boldsymbol Z^T \boldsymbol Z)^{-1})$. 

Now that we have fully specified our model, we turn to the decision-theoretic recipe for identifying differential abundance. After the Gibbs sampling generates a sample from the posterior, we can then compute an MCMC estimate of Pr$(\alpha(A)\neq 0|\boldsymbol Y)$, which is called the {\em posterior marginal alternative probability} (PMAP) for each $A\in\mathcal{I}$, 
and Pr$(\boldsymbol\alpha\neq \boldsymbol 0\,|\,\boldsymbol Y)$, called the {\em posterior joint alternative probability} (PJAP).
One can then use as a decision rule when PJAP is larger than some threshold to reject the {\em global} null hypothesis that no difference at all exists in any microbial taxa between the two groups. The PMAPs, on the other hand, can be used to decide what microbial taxa are called to be significantly different across groups. The PMAPs and the PJAP can all be directly estimated from the posterior samples of $\boldsymbol\alpha$. 

\subsection{Numerical experiments}
\label{sec:numerical_examples}

Before the case study we carry out simulation experiments whose underlying signals are known to assess the behavior of the LTN-based mixed-effects model and compare its performance to other popular methods for differential abundance analysis. To avoid specifying the actual generative mechanisms of the microbiome compositions, we generate synthetic data by injecting signals into the 16S rRNA sequencing data of T1D cohort of DIABIMMUNE project \citep{diabimmune}, which collects longitudinal microbiome samples from infants from Finland and Estonia. The full dataset contains counts of 2240 OTUs of 777 samples from 33 subjects. There are 25 subjects that have more than 20 samples, 7 subjects whose number of samples is in $(10, 20]$, and only 1 subject that has less than 10 samples. 
For illustration, we focus on the 100 OTUs with the highest relative abundance. The phylogenetic tree of the 100 OTUs is used in our LTN-based mixed-effects model. 

We vary the following a triplet parameters that controls the simulation setting---namely, the sample size $n^*$, the effect size per OTU $a^*$, and the number of differentially abundant OTUs $K^*$. Specifically, for each combination of $(n^*,a^*,K^*)$, in a single simulation run, synthetic datasets are generated as follows. For each of the 33 subjects in the DIABIMMUNE cohort, we randomly sample (without replacement) $n^*$ samples for that subject. If there are less than $n^*$ samples for that subject in the dataset, we just draw all of the available samples for that subject. A null dataset is created by randomly dividing the samples into two equal-sized groups, namely, group 0 and group 1. An alternative dataset is generated by randomly choosing $K^*$ OTUs (without replacement) from the 20 OTUs with highest average relative abundance and  multiplying the counts of these $K^*$ OTUs in group 1 by $(1 + a^*)$. For each combination $(n^*, a^*, K^*)$, 500 datasets for the null and for the alternative respectively.

We consider $n^*\in\{10, 20\}$ and $K^*\in\{1, 8\}$. $K^*=1$ and $K^*=8$ correspond to cross-group differences at single and multiple OTUs respectively. 
The single OTU association case ($K^*=1$) is the most favorable scenario to single OTU scanning approaches such as MaAsLin2 \citep{MaAsLin2} and least favorable to methods that jointly model all taxa such as LTN. We expect many common outcomes of interest, especially complex phenotypes, such as the seroconversion status in our case study to be associated with multiple OTUs simultaneously, which we imitate using the $K^*=8$ case. For each combination of $(n^*,K^*)$, we examine the performance of the three methods under weak, medium and strong signal specified in Table~\ref{tab:signal}.

\begin{table}[H]
    \centering
    \begin{tabular}{c|ccc}
    \hline
        & weak & medium & strong \\
         \hline 
         $K^* = 1$ (single OTU) & $a^* = 0.5$ & $a^* = 2$ & $a^* = 4$\\
         $K^* = 8$ (multiple OTUs) & $a^* = 0.5$ & $a^* = 0.75$ & $a^* = 1$\\
         \hline
    \end{tabular}
    \caption{The specification of signal strength under single and multi-OTU associations.}
    \label{tab:signal}
\end{table}

It is worth emphasizing that throughout our experiments the associated single or multiple OTUs are all randomly selected {\em without} regard to the underlying tree structure, and therefore the phylogenetic tree is irrelevant to the underlying outcome of interest, corresponding to the case that the underlying dyadic tree is grossly misspecified. As such the numerical results provides conservative estimates of the performance of LTN and gives evidence on the robustness of our LTN-based model with respect to tree misspecification.

We compare the performance of our LTN-based mixed-effects model in testing the existence of cross-group differences with that of MaAsLin2 \citep{MaAsLin2} and that of DirFactor \citep{dirfactor}. MaAsLin2 tests the differential abundance one OTU at a time, whereas similar to our approach, DirFactor is based on a joint model of the OTU composition, but does not allow the identification of invidual OTUs that are differentially abundant. We apply all three three methods on the same simulated datasets to test the existence of difference in microbiome compositions between the two groups, adjusting for country, age at collection and case/control status of the samples and incorporating subject-level random effects. 

\begin{figure}[p]
    \centering
    \begin{subfigure}[b]{0.8\textwidth}
    \centering
    \includegraphics[width=\textwidth]{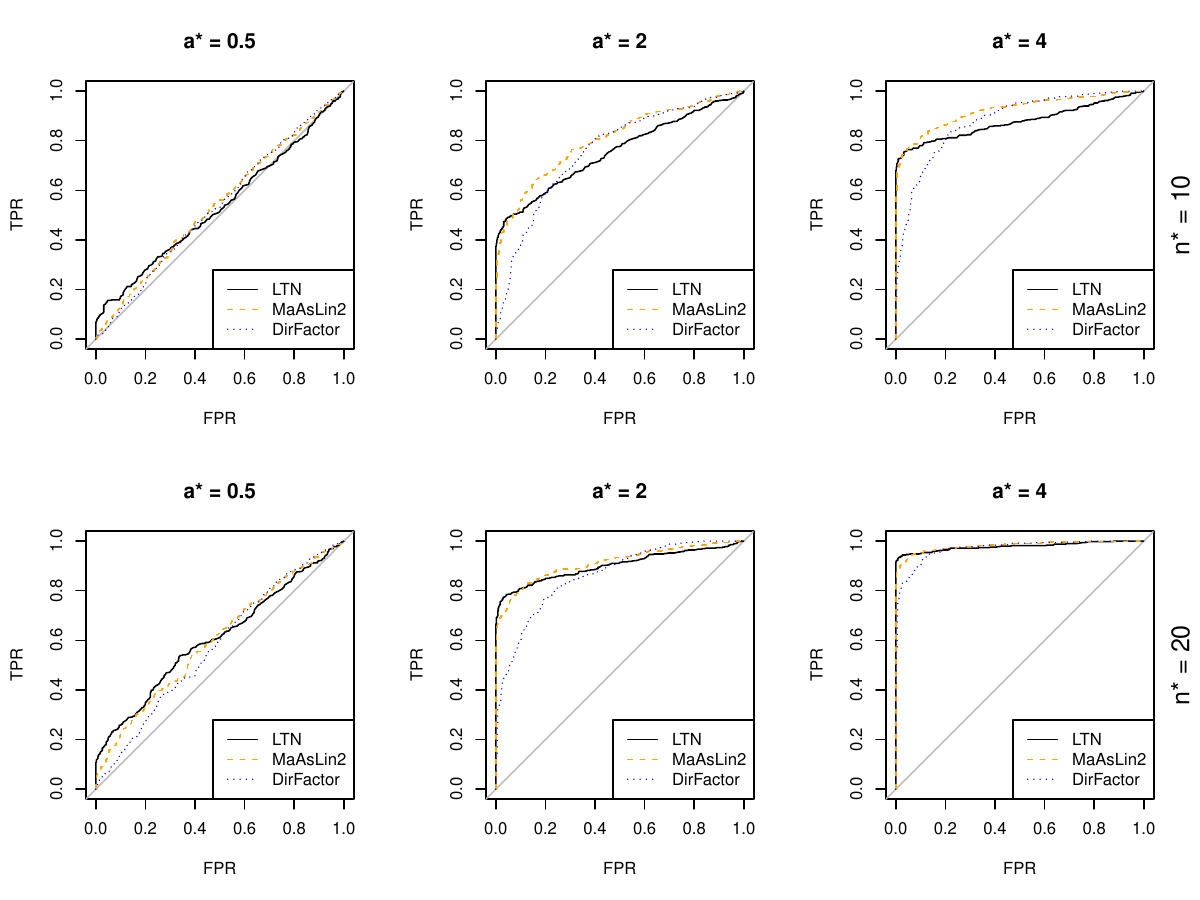}
    \subcaption{Cross-group differences exist at a single OTU.} 
    \end{subfigure}
    \begin{subfigure}[b]{0.8\textwidth}
    \centering
    \includegraphics[width=\textwidth]{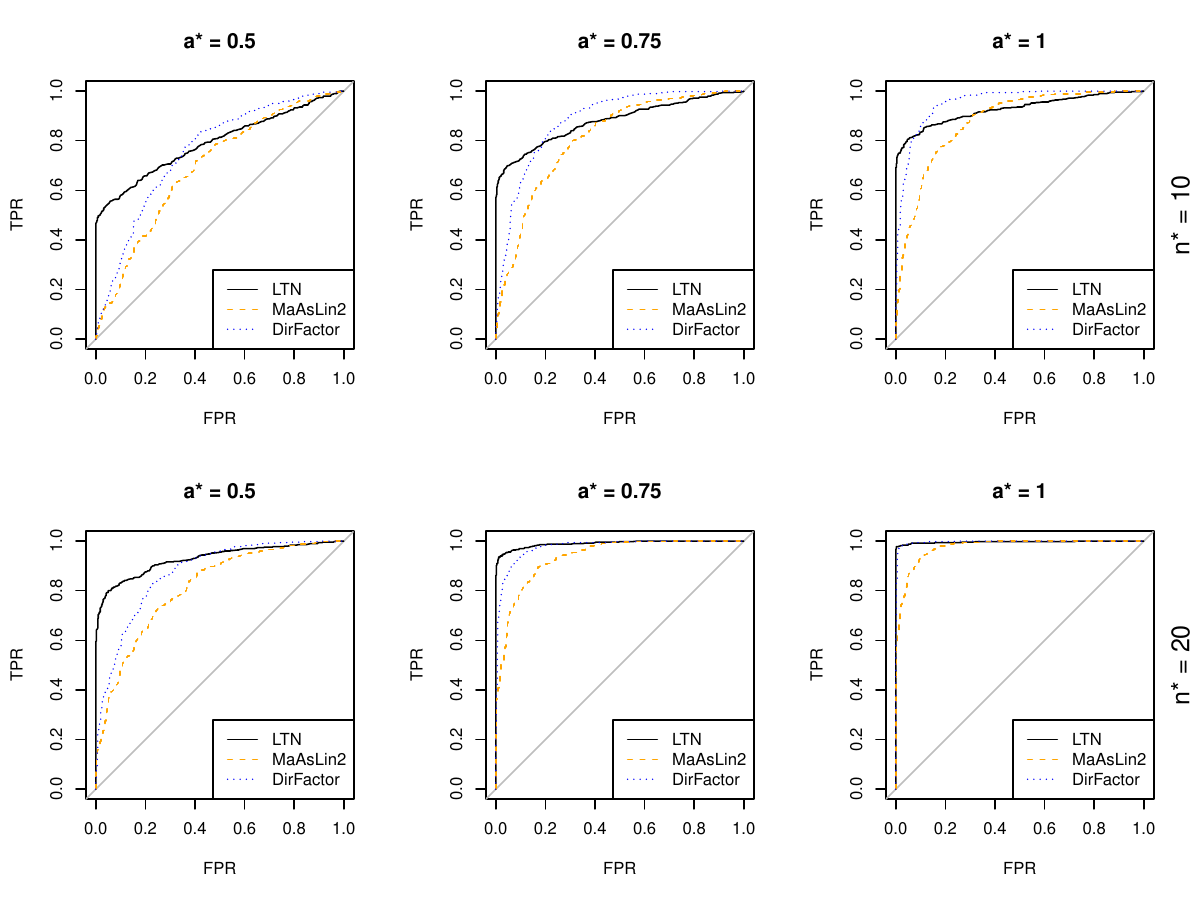}
    \subcaption{Cross-group differences exist at 8 OTUs.}
    \end{subfigure}
    \caption{ROC curves by the signal level. MaAsLin2 is fitted with the R package MaAsLin2 with default parameters, and DirFactor is fitted with the code provided at \url{https://github.com/boyuren158/DirFactor-fix}. The ROC curves were produced using the ROCR package in R \citep{rocr}. }
    \label{fig:roc_matrix}
\end{figure}

The performance of our model, DirFactor and MaAsLin2 are evaluated based on the ROC curves (Fig~\ref{fig:roc_matrix}) for testing the existence of differences in microbiome compositions between the two groups. As defined in Section \ref{sec:mixed_effects}, the (global) test statistic of our model is the posterior joint alternative probability (PJAP). For MaAsLin2, we use the smallest Benjamini-Hochberg (BH) q-values of the associations between the group label (0/1) and the OTUs as the test statistic, and for DirFactor, we use the $l_2$ norm of the coefficient of group label the test statistic as suggested in \cite{dirfactor}. 

The performance of our method is robust for both $K^*=1$ and 8. The $K^*=1$ scenario is most favorable to single-OTU methods such as MaAsLin2, which yields the best ROC curves generally as one would expect. Even in this case, our fully multivariate LTN-based mixed-effects model achieves comparable ROC with sufficient sample size ($n^*=20$) and outperforms DirFactor, which is also a multivariate approach. When cross-group difference occur at multiple OTUs ($K^*=8$), jointly modeling the abundance, as in our method and DirFactor, greatly improves the power over single-OTU testing, and our method outperforms both competitors across all signal levels.

\begin{figure}
    \centering
    \includegraphics[width=0.7\textwidth]{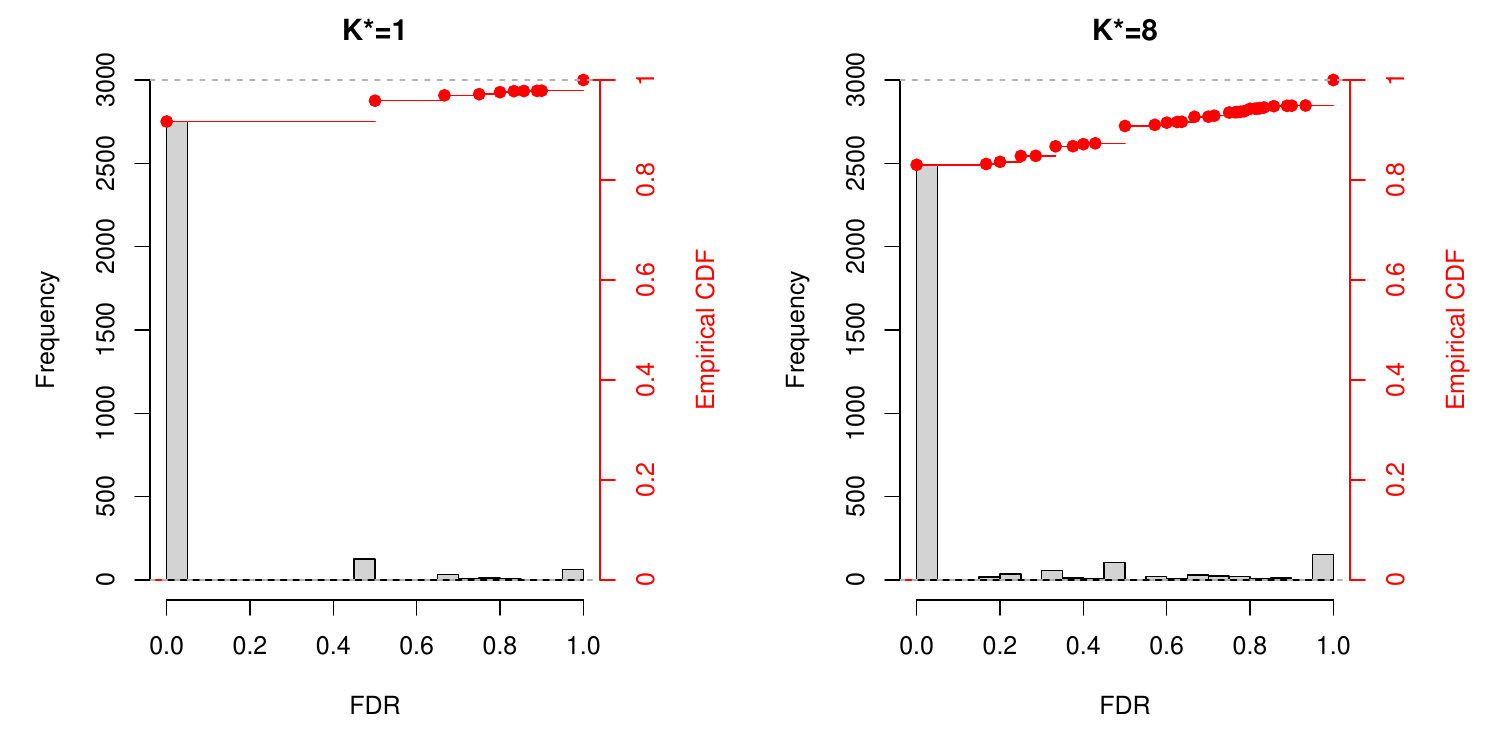}
    \caption{Histogram of the observed FDR for MaAsLin2 in the simulation runs with a target FDR 0.05 for $K^*=1$ and $8$ respectively. For $K^*=8$ in particular, about 20\% of the times the actual FDR was over 0.2 while the normal level is set at 0.05.} 
    \label{fig:fdr_MaAsLin2}
\end{figure}

In addition, we have observed in our simulation settings that the univariate approach MaAsLin2 did not appear to properly control the FDR under the multi-OTU scenario $(K^*=8)$ using BH multiplicity adjustment in the sense that the actual empirical FDR can have very high variance. To see this, Figure \ref{fig:fdr_MaAsLin2} presents the empirical FDR (at a targeted FDR of 5\%) over the simulation runs, and for $K^*=8$, the FDR can sometimes be substantially higher than the target value of 5\%. In fact, for $K^*=8$, more than 10\% of the times, the FDR can be even higher than 50\%, a ten-time inflation of the nominal target FDR. This may later explain the large number of associated OTUs we report later in the case study when applying MaAsLin2. 

\section{Case study: the DIABIMMUNE data}
\label{sec:case-study}
\subsection{Overview of the data and the analysis}
We now carry out a more in-depth analysis of the Type~1 diabetes (T1D) cohort of DIABIMMUNE project to study the relationship of microbiome composition with several variables.  The main goal of this study is to compare microbiome in infants who have developed T1D or serum autoantibodies (markers predicting the onset of T1D) with healthy controls in the same area. Within the time-frame of the study, 11 out of 33 infants seroconverted to serum autoantibody positivity, and of those 11 infants, four developed T1D. Previous studies on this dataset have established associations between the microbiome composition with both the T1D status as well as 
dietary factors. For example, it has been shown that the microbiome of T1D patients tend to have reduced $\alpha$-diversity and increased prevalence of Bacteriodetes taxa \citep{dedrick2020,diabimmune,brugman}. In the following analysis we focus on the seroconversion status and the dietary covariates, and study the relationship between microbiome composition with each of these variables. We again apply our LTN-based mixed-effects model and compare the results to those from MaAsLin2 and DirFactor to the extent possible. 

Key findings from our analysis are summarized as follows: 
(\romannumeral 1) the microbiome compositions of samples taken after seroconversion differ from the other samples in terms of the relative abundance {within Bacteroides, Parabacteroides, and  Erysipelotrichacea, {which are different from the taxa associated with T1D identified in \cite{diabimmune}}}; (\romannumeral 2) introduction of several kinds of food and cessation of breastfeeding can significantly alter the gut microbiome of the infants. We present details of our study on different covariates in the following sections.

\subsection{Differential abundance analysis over seroconversion status}
\label{sec:t1d}
The T1D status of the individuals at the end of the study has been recorded as one of the three levels: control, seroconverted, and T1D at onset. This status is used to define the case and control groups in the original study, where the case group consists of both seroconverted individuals and those who are clinically classified as T1D at onset. For those in the case group, the age at seroconversion is also recorded. Previous analysis on the T1D status reveals that after the ``seroconversion window" (i.e., the time between the first and third quartiles of age at seroconversion for all seroconverted and T1D subjects), the alpha-diversity of microbial communities in T1D subjects showed a significant flattening, while the alpha diversity in other subjects continue to increase, and such difference can be accounted for by several groups of bacteria \citep{diabimmune}. 

Motivated by these findings and inspired by a previous re-analysis on this data given in \cite{dirfactor}, we perform a differential abundance between the samples collected before and after seroconversion to identify the nature of the changes in microbiome compositions after seroconversion. We note that while \cite{dirfactor} also investigated the association of microbiome composition with seroconversion status they focus on {\em estimating} the covariate effects on the individual taxa without providing a formal means to {\em testing} on differential abundance of the taxa.
Following \cite{diabimmune}, we also include eight binary dietary variables that indicate whether a specific type of diet was consumed at the time when the sample was collected: \textit{breastfeeding, solid food, eggs, fish, soy products, rye, barley}, and \textit{buckwheat and millet}. Dietary patterns are known to influence microbiome composition significantly and therefore must be accounted for. Age at collection (log-transformed), gender and nationality of the individuals are also included in our analysis as covariates.

\begin{figure}[thb]
    \centering
    \includegraphics[width=0.9\linewidth]{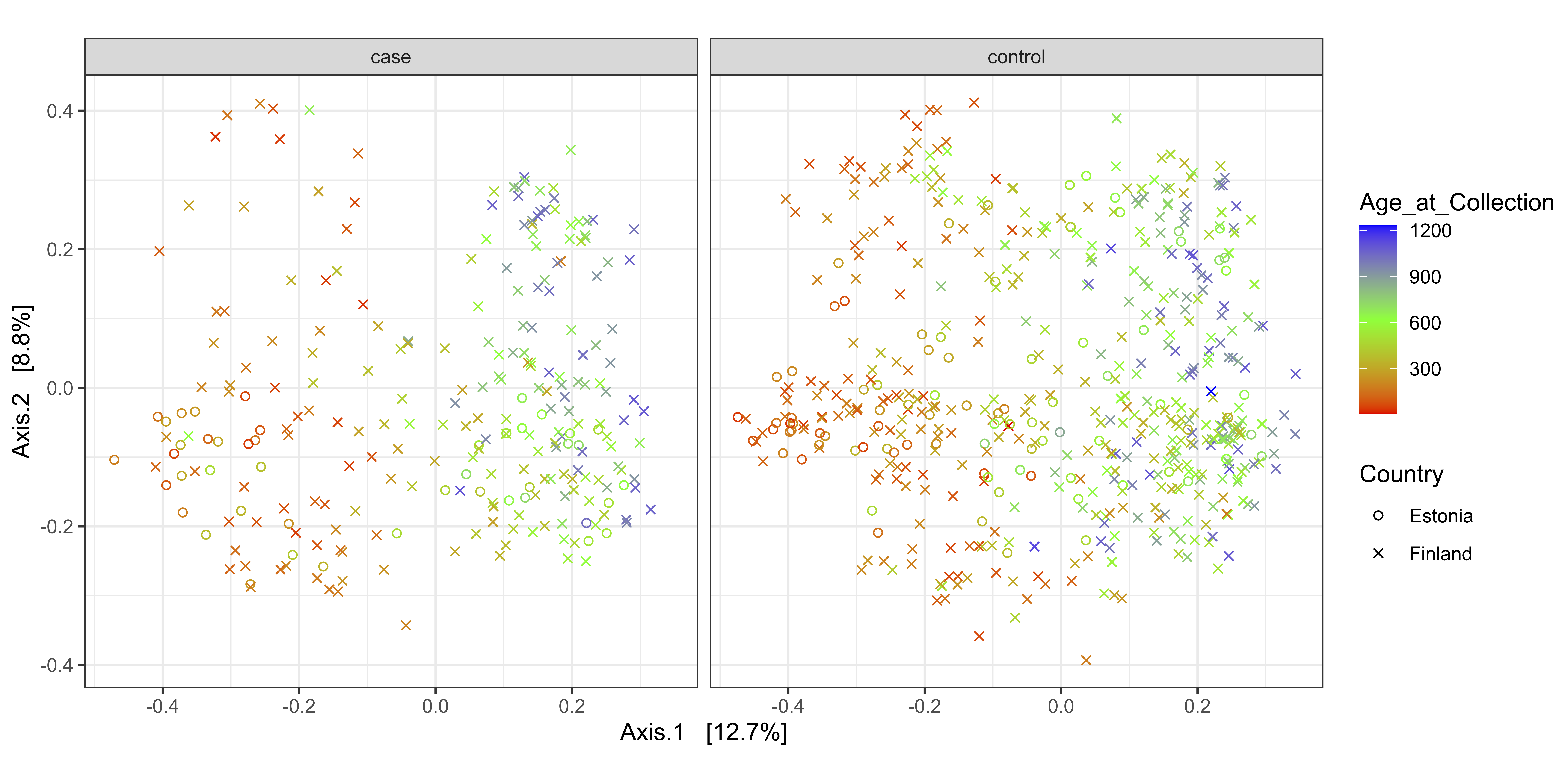}
    \caption{Multidimensional scaling of Bray-Curtis dissimilarity between samples}
    \label{fig:mds}
\end{figure}

\begin{figure}[thb]
    \centering
    \includegraphics[width=0.9\linewidth]{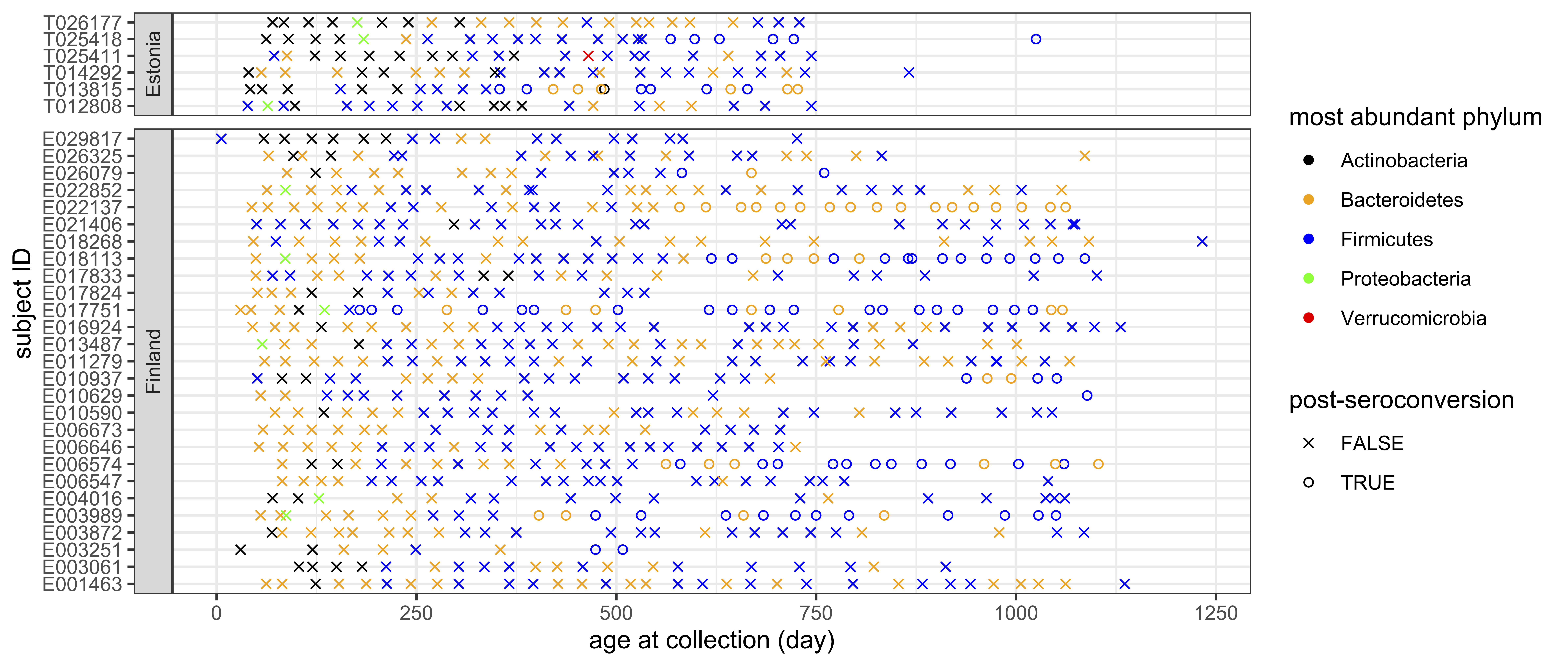}
    \caption{The most abundant phylum in each sample.}
    \label{fig:most_abundant_phylum}
\end{figure}

{\em Exploratory analysis.} Figure~\ref{fig:mds} shows a multidimensional scaling plot using Bray-Curtis dissimilarity between samples, where age at collection increases almost in the same direction of the first axis, and the Finish and Estonian samples are roughly separated along the second axis in the case group. There is a clear general trend in age and large cross-country difference, as well as substantial variability across subjects from the same country at similar ages. Figure~\ref{fig:most_abundant_phylum} presents the most abundant phylum in each subject over time and it shows large variability at birth both between subjects and within, which gradually stabilizes over time. As such, we use log(age) as a covariate to account for the more rapid changes in microbiome compositions at early age. Interestingly, seroconversion generally happens after the highly variable phase. Most samples collected after one year old are dominated with either Bacteroidetes or Firmicutes. Moreover, Actinobacteria dominates other taxa in about half of the samples collected from Estonian individuals  during the first year, while that is not the case for Finnish individuals.

{\em Model-based differential abundance analysis.} We next fit our LTN-based mixed-effects model to identify differences in OTU compositions over seroconversion status and focused on the top 100 OTUs with highest relative abundance. Specifically, sample $i$ has group label $s_i=1$ if it is collected after seroconversion and $s_i=0$ otherwise. We include environmental and diet factors that are deemed likely to be associated with the microbiome composition as fixed effects. In addition, individual random effects are included to account for the between-individual variation. We fit MaAsLin2 and DirFactor with the same set of fixed and random effects for comparison.  

For the purpose of testing the global null, we set $m=1-0.5^{1/99}\approx 0.006977$, corresponding to a prior probability of 0.5 for the global null to hold. The corresponding PJAP is close to 1, indicating overwhelmingly strong evidence against null, which is unsurprising and consistent with the results from  MaAsLin2 and DirFactor, both of which result in a permutation p-value $<0.001$ under 1,000 permutations. For the more interesting task of identifying specific taxa that are differentially abundant, we set $m=0.05$, which corresponds to a prior expected proportion of 5\% for nodes on which there is differential splitting probabilities. This more relaxed multiple testing adjustment strategy fall more in line with the practice of setting prior expected number of differentially abundant taxa. DirFactor does not in fact allow direct evaluation of individual taxon-level association testing so are excluded in the comparison. 

\begin{figure}[H]
    \centering
    \includegraphics[width=0.5\textwidth,angle=90]{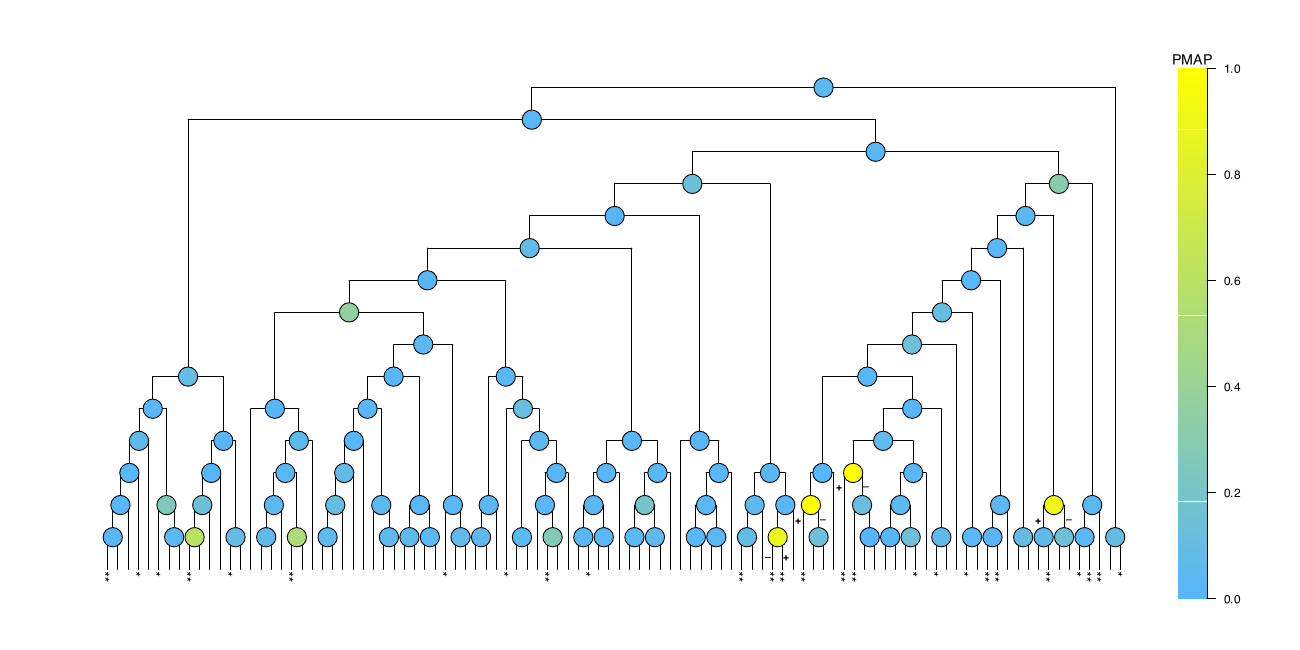}
    \caption{Posterior marginal alternative probabilities (PMAPs) of nodes from LTN-based mixed-effects model and BH q-values of OTUs from MaAsLin2 in the comparison between pre- and post-seroconversion samples. 
    The OTUs with BH q-value reported by MaAsLin2 $\leq 0.05$ and $\leq 0.1$ are marked with ``**" and ``*" respectively. The ``+" and ``-" on branches are labelled according to the sign of posterior mean of $\alpha(A)$'s, indicating whether the left or the right child nodes has increased relative abundance among the cases. 
    }
    \label{fig:pmap_qval_m005_seroconversion}
\end{figure}

Figure \ref{fig:pmap_qval_m005_seroconversion} presents the estimated PMAPs from our method for all the interior nodes of the phylogenetic tree. For comparison, in the same figure we label the leaf nodes (i.e., the OTUs) with BH-adjusted q-values ($\leq 0.05$ and $\leq 0.1$) reported by MaAsLin2. The four nodes with the highest PMAPs suggest that changes occur in relative abundance of OTUs within the genus Bacteroides, the family Erysipelotrichaceae, and the genus Parabacteroides around the time of seroconversion. In particular, the results from our method further suggests increased abundance of P. distasonis relative to its sibling on the phylogenetic tree after seroconversion, which is consistent with the observed relative abundance of P. distasonis in the data (see Figure \ref{fig:distasonis}). These identified associations are supported by previous studies based on other datasets \citep{microorganisms9071436Bacteroides, mejia2014fecaldistasonis}. We note that while MaAsLin2 reports a much larger set of significant OTUs associated with the seroconversion level, we think these could contain a fairly large number of false positives given the inflated empirical FDR in our simulation study for the multi-OTU association case (as shown in Figure~\ref{fig:fdr_MaAsLin2}).

\begin{figure}[H]
     \centering
     \begin{subfigure}[b]{0.45\textwidth}
         \centering
         \includegraphics[width=\textwidth]{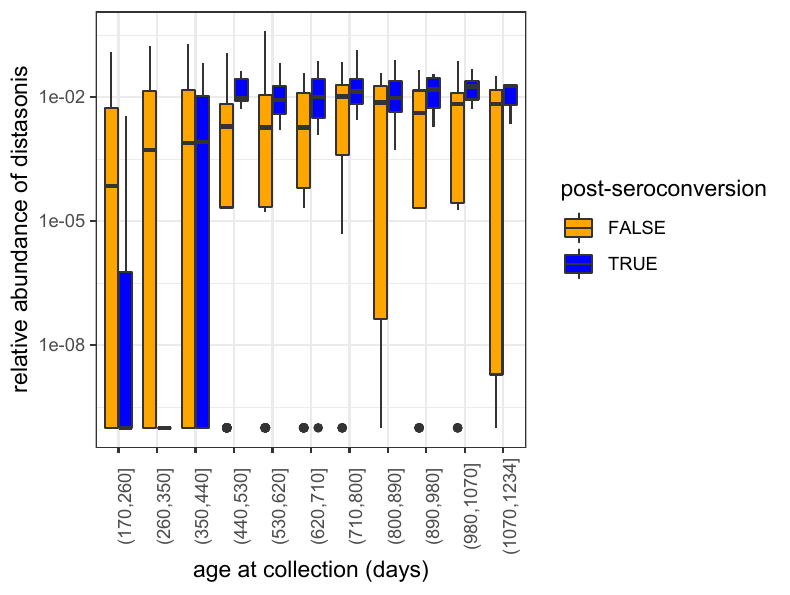}
         \caption{P.\ distasonis}
         \label{fig:distasonis}
     \end{subfigure}
     \hfill
     \begin{subfigure}[b]{0.45\textwidth}
         \centering
         \includegraphics[width=\textwidth]{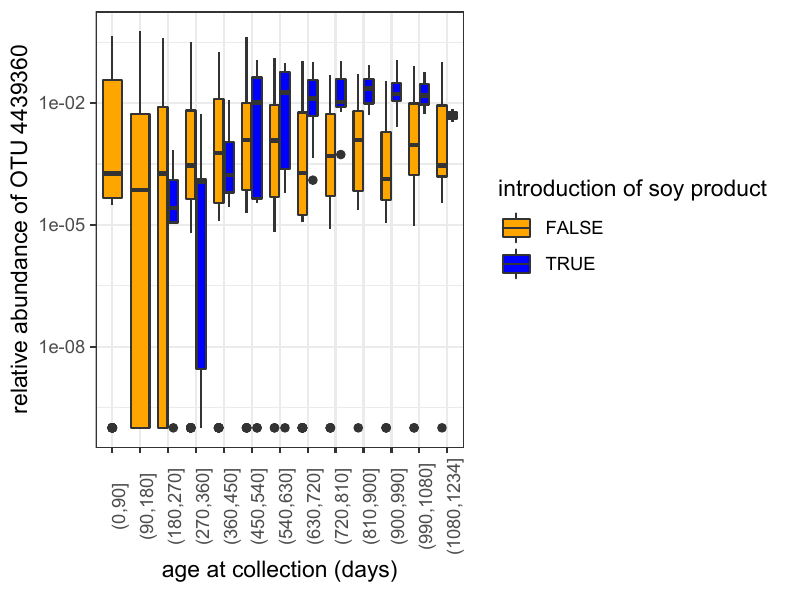}
         \caption{OTU 4439360} \label{fig:4439360}
     \end{subfigure}
        \caption{Relative abundance of P.\ distasonis and OTU 4439360. For P. distasonis, data collected before the first event of seroconversion is omitted.}
\end{figure}

\subsection{Goodness-of-fit to the within-OTU and within-sample zeros}
\begin{figure}[H]
     \centering
     \begin{subfigure}[b]{0.4\textwidth}
         \centering
         \includegraphics[width=\textwidth]{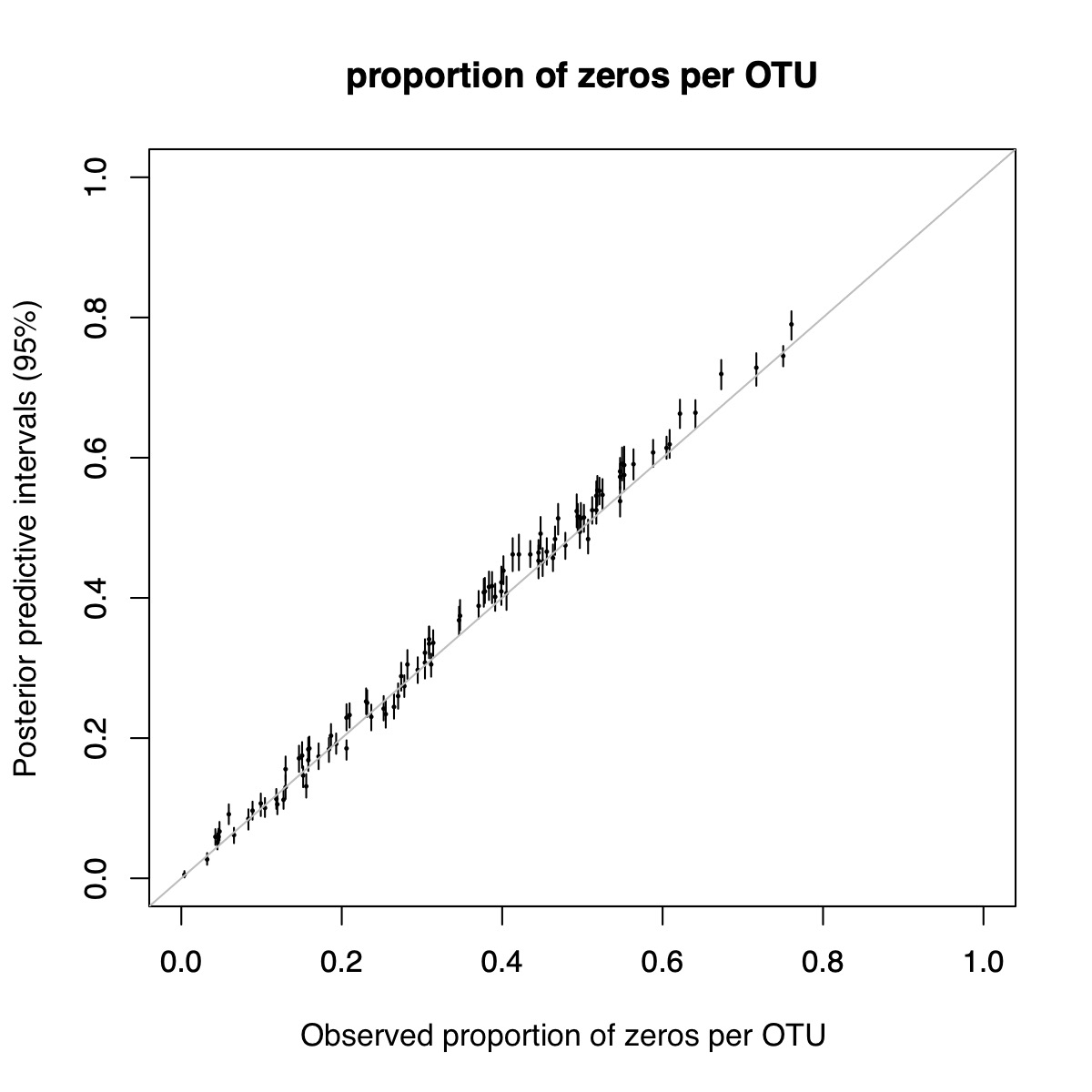}
         \caption{Proportion of zeros per OTU (observed vs posterior predictive). The bars represent 95\% posterior predictive intervals of the proportion of zeros per OTU (calculated based on 500 posterior predictive samples).}
         \label{fig:otu_zero_diabimmune}
     \end{subfigure}
     \hfill
     \begin{subfigure}[b]{0.4\textwidth}
         \centering
         \includegraphics[width=\textwidth]{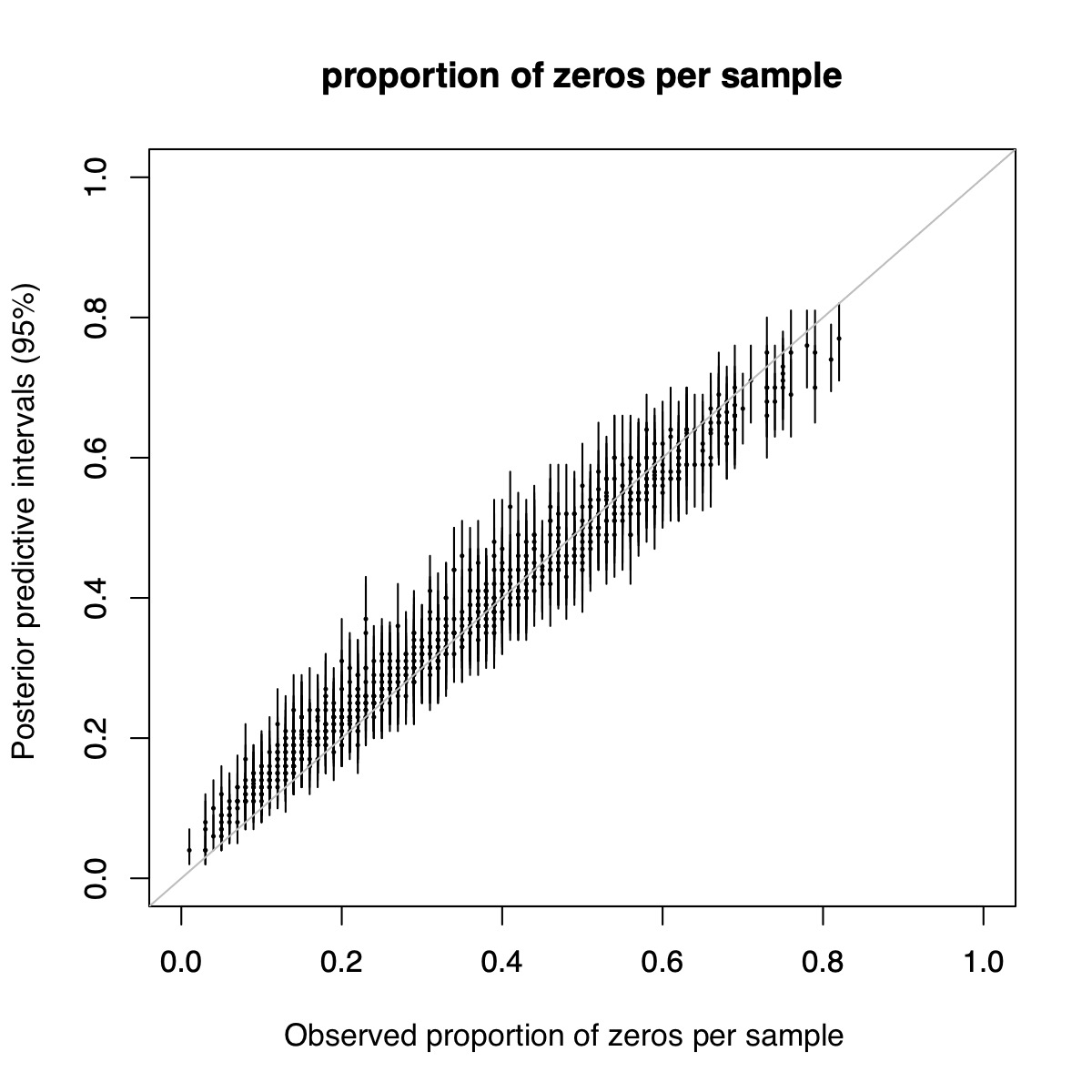}
         \caption{Proportion of zeros per sample (observed vs posterior predictive). The bars represent 95\% posterior predictive intervals of the proportion of zeros per sample (calculated based on 500 posterior predictive samples). }
         \label{fig:sample_zero_diabimmune}
     \end{subfigure}
        \caption{Posterior predictive intervals for proportions of zeros by OTU (a) and by sample (b).}
        \label{fig:zeros_diabimmune}
\end{figure}

To examine how adequately the zero patterns in the data are captured by our proposed model, we fit our LTN mixed-effects model with a matrix normal prior on the coefficients of all covariates (i.e., without a spike-and-slab component) to the DIABIMMUNE data, and generate posterior predictive samples of OTU counts. The posterior predictive distribution of the proportion of zeros is shown in Figure~\ref{fig:zeros_diabimmune}. In general, our model is capable of capturing both within-OTU and within-sample sparsity in this dataset without explicitly modeling a zero-inflation component, which is consistent with the arguments given in  \cite{silverman2020_zero} and \cite{sarkar_stephens2021} that adequately modeling cross-sample heterogeneity under a count sampling model is often adequate in characterizing the abundant zeros in sequencing data.

\subsection{Differential abundance analysis over dietary variables}
As additional examples that showcase the performance of our method and to compare it to MaAsLin2, we repeat the same abundance analysis on the microbiome compositions to each of the eight dietary factors. Due to space constraints, we report all of the results in Supplementary Materials~\ref{sec:dietary}.

\section{Conclusion}
We have introduced a tree-based generative model called LTN for microbiome compositions. LTN decomposes the multinomial sampling model to a collection of binomials at the interior nodes of the phylogenetic tree, and  transforms OTU compositions to node-specific log-odds and adopts multivariate normal model on the log-odds, thereby offering a flexible covariance structure among OTUs. With the tree-based decomposition, LTN avoids the computational challenges caused by the lack of conjugacy between multinomial and multivariate normal in LN models, and efficient inference can be carried out with Gibbs sampling by introducing P\'olya-Gamma auxiliary variables. This opens the door to adopting a wide range of multivariate analysis models/methods for microbiome analysis while maintaining computational tractability.

Due to its fully generative nature, LTN can be used either as a standalone model or a component in more sophisticated models such as those involving covariate effects and latent structures, which are often needed in microbiome applications. 
We have provided examples of mixed-effects modeling for cross-group comparison. 
We have shown how to carry out cross-group comparison of microbiome compositions through a Bayesian model choice framework. In these examples, we had adopted the gLasso prior for demonstration but our model/prior choices can and should be substituted as the practitioner sees fit in their given context. Finally, 
beyond compositional data, tree-based log-odds decompositions can also be used as device for modeling general probability distributions \citep{jara&hanson:2010}, the modeling and inference strategies described herein can be applied in these broader contexts as well.

\section*{Software}
We provide an R package ({\tt LTN}: \url{https://github.com/MaStatLab/LTN}) implementing the LTN-based mixed-effects model. 

\section*{Acknowledgements} 
This research is partly supported by NIH grant R01-GM135440 as well as NSF grants DMS-1749789 and DMS-2013930.

\bibliography{arxiv}

\newpage
\resetlinenumber
\beginsupplement

\section*{Supplementary materials}

\renewcommand{\thesubsection}{\Alph{subsection}}
\renewcommand{\thefigure}{S\arabic{figure}}
\renewcommand{\theequation}{S\arabic{equation}}

\subsection{Technical details for Bayesian computation with LTN}
\subsubsection{P\'olya Gamma augmentation for LTN}\label{sec:s-pg}
In this section we describe the P\'olya Gamma augmentation scheme for the proposed LTN method. The binomial sampling model is
$$y(A_l)| \psi(A),y(A)) \ind {\rm Binomial}\left(y(A),\frac{e^{\psi(A)}}{e^{\psi(A)}+1}\right)\qquad\text{ for all } A \in \mathcal{I}$$
Following \cite{polson}, we can write the binomial likelihood for a sample at an interior node $A$ as
\begin{align*}
 p(y(A_l)|y(A),\psi(A))\propto   \frac{\left(e^{\psi(A)}\right)^{y(A_l)}}{\left(1+e^{\psi(A)}\right)^{y(A)}} &= 2^{-y(A)}e^{\kappa(A)\psi(A)}\int_0^\infty e^{-w \psi(A)^2/2}f(w)dw,
\end{align*}
where $\kappa(A)=y(A_l)-y(A)/2$ and
\begin{equation*}
 f(w)=\frac{2^{y(A)-1}}{\Gamma(y(A))} \sum_{n=0}^{\infty}(-1)^{n} \frac{\Gamma(n+y(A))}{\Gamma(n+1)} \frac{(2 n+y(A))}{\sqrt{2 \pi w^{3}}} e^{-\frac{(2 n+y(A))^{2}}{8 w}}
\end{equation*}
is the probability density function of the P\'{o}lya-Gamma distribution PG$(y(A),0)$. We can accordingly introduce an auxiliary variable $w(A)$ that is independent of $y(A_l)$ given $y(A)$ and $\psi(A)$, with 
$$p(w(A)|y(A),\psi(A))\propto e^{-w(A)\psi(A)^2/2}f(w(A)).$$
In other words, we add the auxiliary variable $w(A)$ into the LTN model in Eq.~\eqref{eq:generate} with
\[ w(A)|y(A),\psi(A) \sim {\rm PG}(y(A),\psi(A)).\] 
The joint conditional distribution for $w(A)$ and $y(A_l)$ given $y(A)$ and $\psi(A)$ is then
$$p(w(A),y(A_l)|y(A),\psi(A)) \propto 
2^{-y(A)}e^{\kappa(A)\psi(A)-w(A)\psi(A)^2/2}f(w(A)),$$
which is a log-quadratic function of $\psi(A)$, and thus is conjugate to the multivariate Gaussian likelihood of $\boldsymbol\psi=\{\psi(A):A\in\mathcal{I}\}$.

\subsubsection{Gibbs sampler for mixed-effects model in section~\ref{sec:mixed_effects}}
\label{sec:s-sampler}
For notational simplicity, we rewrite the model in the following form:
$$\underset{n\times d}{\boldsymbol \Psi}=\underset{n\times 1}{\boldsymbol s}\underset{1\times d}{\boldsymbol\alpha}+\underset{n\times q}{\boldsymbol  Z}\underset{q\times d}{\boldsymbol\beta}+\underset{n\times G}{\boldsymbol H}\underset{G\times d}{\boldsymbol  \Gamma}+\underset{n\times d}{\boldsymbol \epsilon},$$
where 
$H_{ij}=I(g_i=j)$. We denote the P\'olya-Gamma auxiliary variable of node $A$ and sample $i$ with $w_i(A)$, and let $ \Sigma_\epsilon={\rm diag}(\sigma_\epsilon^2(A):A\in\mathcal{I}), \phi_\epsilon(A)=\sigma^{-2}_\epsilon(A)\sim {\rm Gamma}(c_0,d_0)$. 

The sampler cycles through the following steps:
\begin{itemize}
    \item Sample $\boldsymbol \beta$ from
    \begin{align*}
     \boldsymbol\beta|-\sim N_{q\times d} ((\boldsymbol Z^T{\boldsymbol Z})^{-1}{\boldsymbol Z}^T(\boldsymbol\Psi-\boldsymbol H\boldsymbol\Gamma-{\boldsymbol{s}}\boldsymbol{\alpha})(\boldsymbol\Sigma_\epsilon^{-1}+\boldsymbol I/(cn))^{-1}\boldsymbol\Sigma_\epsilon^{-1},\\(\boldsymbol\Sigma_\epsilon^{-1}+\boldsymbol I/(cn))^{-1}\otimes ({\boldsymbol{Z}}^T{\boldsymbol{Z}})^{-1}).
    \end{align*}
    \item For each $A\in \mathcal{I}$, sample $\alpha(A)$ from
    $$\alpha(A)|-\sim (1-\pi'(A))\delta_0(\alpha(A))+\pi'(A)(1-\delta_0(\alpha(A)))N(b(A)s_\alpha^2(A),s_\alpha^2(A)),$$
    where
    $$\pi'(A)=\frac{\pi(A)s_\alpha(A)\phi_\alpha^{1/2}\exp\{b^2(A)s_\alpha^2(A)/2\}}{1-\pi(A)+\pi(A)s_\alpha(A)\phi_\alpha^{1/2}\exp\{b^2(A)s_\alpha^2(A)/2\}},$$
    $$s_\alpha(A)=(\phi_\alpha+\sum_{i=1}^N {s}_{i}^2\phi_{\epsilon}(A))^{-1/2},b(A)=\phi_{\epsilon}(A)\sum_{i=1}^N({s}_{i}(\psi_{i}(A)-\gamma_{i}(A)-\boldsymbol z_i^T\boldsymbol{\beta}(A))).$$
    \item For each $A\in\mathcal{I}$, sample $\pi(A)$ from 
$$\pi(A)|-\sim {\rm Beta}(m+1-I(\alpha(A)=0),1-m+I(\alpha(A)=0)).$$
  \item For $l=1,\cdots,G$,
  let $$n_l=\sum_{i=1}^n I(g_i=l),\boldsymbol C_l=(n_l\boldsymbol\Sigma_\epsilon^{-1}+\boldsymbol\Omega)^{-1},\boldsymbol m_l=\boldsymbol C_l(\boldsymbol\Sigma_\epsilon^{-1}\sum_{i:g_i=l}(\boldsymbol\psi_i-\boldsymbol\beta^T\boldsymbol z_i-s_i\boldsymbol\alpha^T )),$$
  and sample $\boldsymbol\gamma_l$ from
 $$\boldsymbol\gamma_l|-\sim N(\boldsymbol m_l,\boldsymbol C_l).$$
  \item For each $A\in\mathcal{I}$, sample  $\phi_{\epsilon}(A)$ from
  $$\phi_{\epsilon}(A)|-\sim {\rm Gamma}(c_0+n/2,d_0+\sum_{i=1}^n \epsilon_{i}^2(A)/2)$$
  \item For $i=1,\cdots,n$, let $$\boldsymbol{C}_i=({\rm diag}(\boldsymbol w_i)+\boldsymbol\Sigma_\epsilon^{-1})^{-1},\boldsymbol m_i=\boldsymbol C_i(\boldsymbol\kappa_i+\boldsymbol\Sigma_\epsilon^{-1}(\boldsymbol\beta^T\boldsymbol z_i+\boldsymbol\gamma_{g_i}+s_i\boldsymbol\alpha^T)),$$ 
  then sample $\boldsymbol \psi_i$ from
  $$\boldsymbol \psi_i|-\sim N(\boldsymbol m_i,\boldsymbol C_i)$$ 

\item For $i=1,\cdots , n$ and for each $A\in\mathcal{I}$, sample the P\'olya-Gamma variable $w_i(A)$ from 
  $$w_i(A)|-\sim{\rm PG}(y_i(A),\psi_i(A))$$
\item Sample $\phi_\alpha$ from
  $$\phi_{\alpha}|- \sim {\rm Gamma}(t+\sum_{A\in\mathcal{I}}I({\alpha(A)}\neq 0)/2,u+\sum_{A\in\mathcal{I}}{{\alpha}^2(A)}/2)$$
 \item Update $\boldsymbol \Omega$ with the block Gibbs sampling procedure described in Algorithm~1. For the data-augmented target distribution, set $\boldsymbol S=\boldsymbol\Gamma^T\boldsymbol\Gamma$.  
\end{itemize}

\subsection{Sensitivity analysis on tree misspecification}
\label{sec:s-treesensitivity}

We investigate the robustness of the inference of mean and covariance structure to the choice of partition tree. 
Let ${\mathcal{T}_1}$ be a phylogenetic tree such that all of the right children of the internal nodes are leaves, ${\mathcal{T}_2}$ a balanced tree where the left and right subtree of each node have identical shape. 
To differentiate LTNs corresponding to different trees, we use ${\rm LTN}_{\mathcal{T}}$ to indicate a model constructed on tree $\mathcal{T}$.
The leaves of the two trees have the same order in the pre-order traversal trace. The data $\boldsymbol{X}_{n\times K}$ is generated from $\rm{LTN}_{{\mathcal{T}_1}}(\boldsymbol{\mu}_1,\boldsymbol{\Sigma}_1)$, where number of samples $n=200$, number of OTUs $K=64$, $\boldsymbol\mu_1=(2,\cdots,2),\boldsymbol\Sigma_1= \boldsymbol I$. 

We fit ${\rm LTN}_{\mathcal{T}_1}$ and ${\rm LTN}_{\mathcal{T}_2}$ on $\boldsymbol{X}$ to estimate the mean and covariance of the log-odds at the internal nodes of $\mathcal{T}_1$ (ground truth) and $\mathcal{T}_2$ (misspecified tree) respectively, and convert such estimates from the misspecified tree to the correct tree $\mathcal{T}_1$. 

\begin{figure}[h!]
    \centering
    \includegraphics[width=0.8\linewidth]{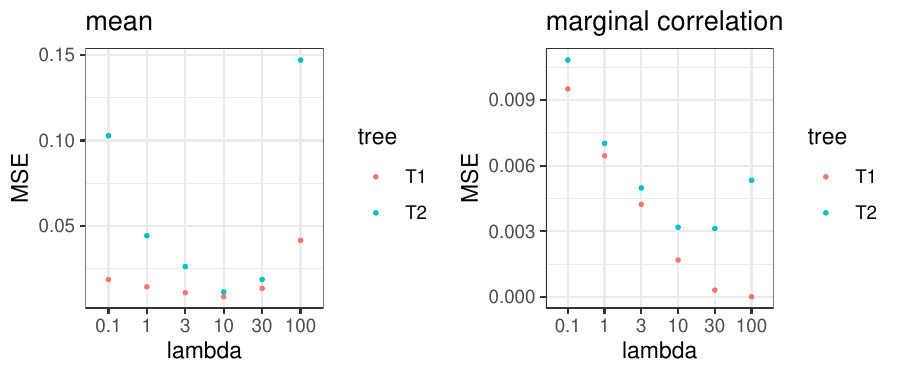}
    \caption{MSE of the estimated mean ($\boldsymbol\mu$) and marginal correlations on the original tree, averaged across all nodes. The MSE is calculated based on 100 replicates. }
    \label{fig:mse_tree}
\end{figure}

We ran 100 replicates under this simulation setting. The MSE of $\bmu$ and marginal correlations on the original tree $\mathcal{T}_1$ averaged over the 63 internal nodes are shown in Figure \ref{fig:mse_tree}. As expected, the estimates under the correct tree almost always have the smallest MSE.  Inference under the misspecified tree seem to be more sensitive to the hyperparameter $\lambda$ than the correct tree.     

\begin{figure}[h!]
    \centering
    \includegraphics[width=0.9\linewidth]{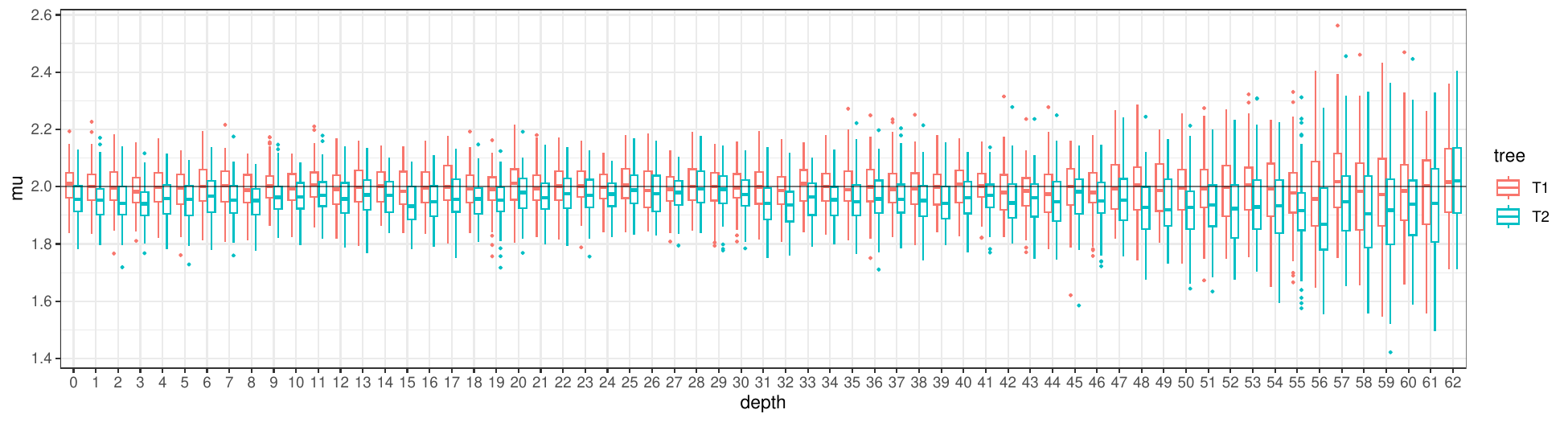}
    \caption{Estimated $\mu$ across nodes with $\lambda = 10$. The boxplots are generated based on 100 replicates. The nodes are ordered by their depth in $\mathcal{T}_1$.}
    \label{fig:boxplot_mu}
\end{figure}

\begin{figure}[h!]
    \centering
    \includegraphics[width=\linewidth]{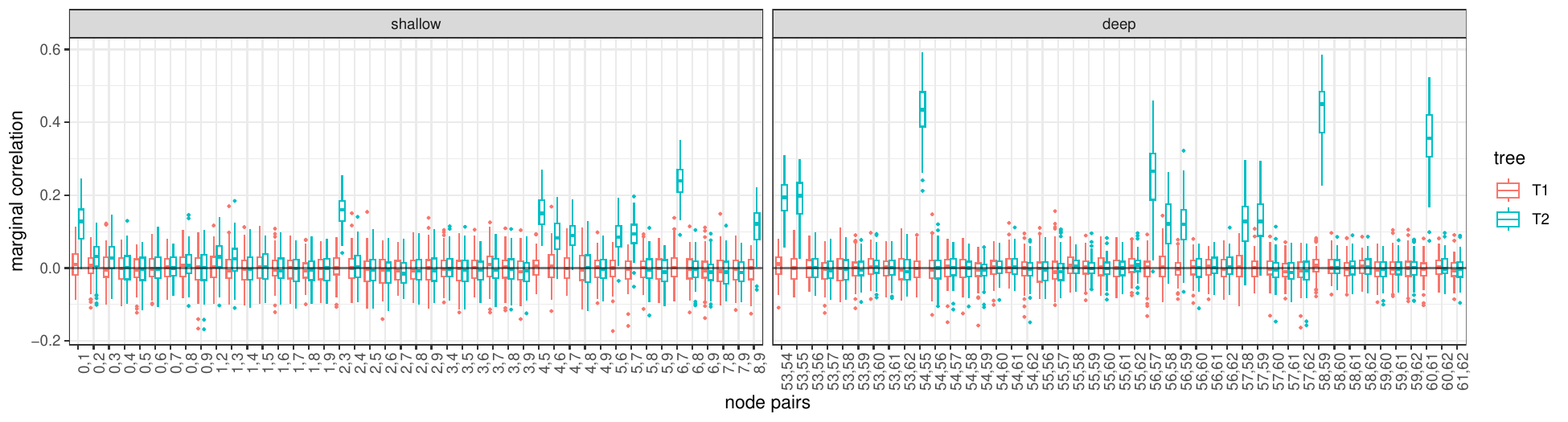}
    \caption{Estimated marginal correlations across nodes with $\lambda=10$. The left panel shows the shallowest ten nodes in $\mathcal{T}_1$ while the right shows the deepest ten nodes. The boxplots are generated based on 100 replicates. The nodes are labeled by their depth in $\mathcal{T}_1$. }
    \label{fig:boxplot_corr}
\end{figure}

Figures \ref{fig:boxplot_mu} and \ref{fig:boxplot_corr} provide a closer inspection of the nodes. The estimated mean and marginal correlations of the shallow nodes are generally robust to the misspecified trees though for a small number of node-pairs, the estimated correlation can be some-off biased. The extent of such bias is generally small for nodes in shallow levels of the tree. We emphasize that here our $\mathcal{T}_2$ is completely randomly generated. In practice, the phylogenetic and taxonomic trees are usually resembling the underlying functional tree to various extents. So the results suggest that in practice inference under LTN is generally robust to misspecification of the tree.

\subsection{Details on the decision-theoretic basis for hypothesis testing based on PMAPs}\label{sec:s-fdr}

The decision on reporting the significant nodes is a multiple testing problem. Let $\boldsymbol y$ be the observed data. Let $d_A(\boldsymbol y) \in \{0,1\}$ be the decision rule on node $A$, where $d_A(\boldsymbol y) = 1$ if and only if the node-specific null $H_0(A): \alpha(A) = 0$ is rejected. Then the posterior expected false discovery rate is 
\begin{align*}
     \overline{\rm FDR}=&E(\frac{\sum_A 1(H_0(A)\text{ is true})d_A(\boldsymbol y)}{\sum_Ad_A(\boldsymbol y)}|\boldsymbol y)\\
    =&\frac{\sum_A P(\alpha(A)=0|\boldsymbol y)d_A(\boldsymbol y)}{\sum_A d_A(\boldsymbol y)}\\
    =&\frac{\sum_A d_A(\boldsymbol y)(1-{\rm PMAP}(A))}{\sum_A d_A(\boldsymbol y)}
\end{align*}

This is the average of $(1-{\rm PMAP})$ over all nodes that are reported as significant. Here we adopt the decision rule $d_A(\boldsymbol y) = 1({\rm PMAP}(A) \geq p)$ for some $p$. To control $\overline{\rm FDR} \leq c$, we sort the $(K-1)$ PMAPs as $p_1\geq p_2\geq \cdots \geq p_{K-1}$, and set the threshold on PMAP in the decision rule as $p = p_k$, where
$$k = \max \{k:k\in \{1,\cdots, K-1\}\text{ and }\frac{1}{k}\sum_{i=1}^k (1-p_i) \leq c\} .$$

\subsection{Details on the differential abundance analysis over dietary variables in the case study}
\label{sec:dietary}
We repeat the same abundance analysis on the microbiome compositions to each of the eight dietary factors. For each diatery variable, we treat all the other variables including the seroconversion status as covariates. Because the cessation of breastfeeding and the introduction of other types of food are registered continuously, we are essentially comparing the microbiome compositions in samples taken before and after those change points of dietary patterns. Similar to the analysis on seroconversion status, in each comparison, we include the T1D status, other dietary covariates, and environmental covariates including age, gender and nationality as fixed effects, individuals as random effects in our mixed effects model and MaAsLin2. The sample size under different comparisons and the PJAPs (with $m=1-0.5^{1/99}$) are shown in Table~\ref{table:pjap}.

\begin{table}[h]
\centering
\caption{PJAPs of the dietary variables}
\label{table:pjap}
\begin{tabular}{cccc}
  \hline
   Variable&
  \multicolumn{2}{c}{Group size}& PJAP\\ 
  \hline 
    & On ($s_i=1$) & Off ($s_i=0$) &\\
    \cline{2-3}
   Barley & 531 & 246 & 1.00\\ 
        Breastfeeding & 248 & 529 & 1.00 \\ 
        Buckwheat\&Millet & 198 & 579 & 1.00 \\ 
        Eggs & 477 & 300 & 0.75 \\ 
        Fish & 581 & 196 & 1.00 \\ 
        Rye & 514 & 263 & 0.97 \\ 
        Solid Food & 681 & 96 & 1.00 \\
        Soy Product & 107 & 670 & 1.00
\\\hline
\end{tabular}
\\
\end{table}

\begin{figure}[t]
    \centering
    \includegraphics[width=\textwidth]{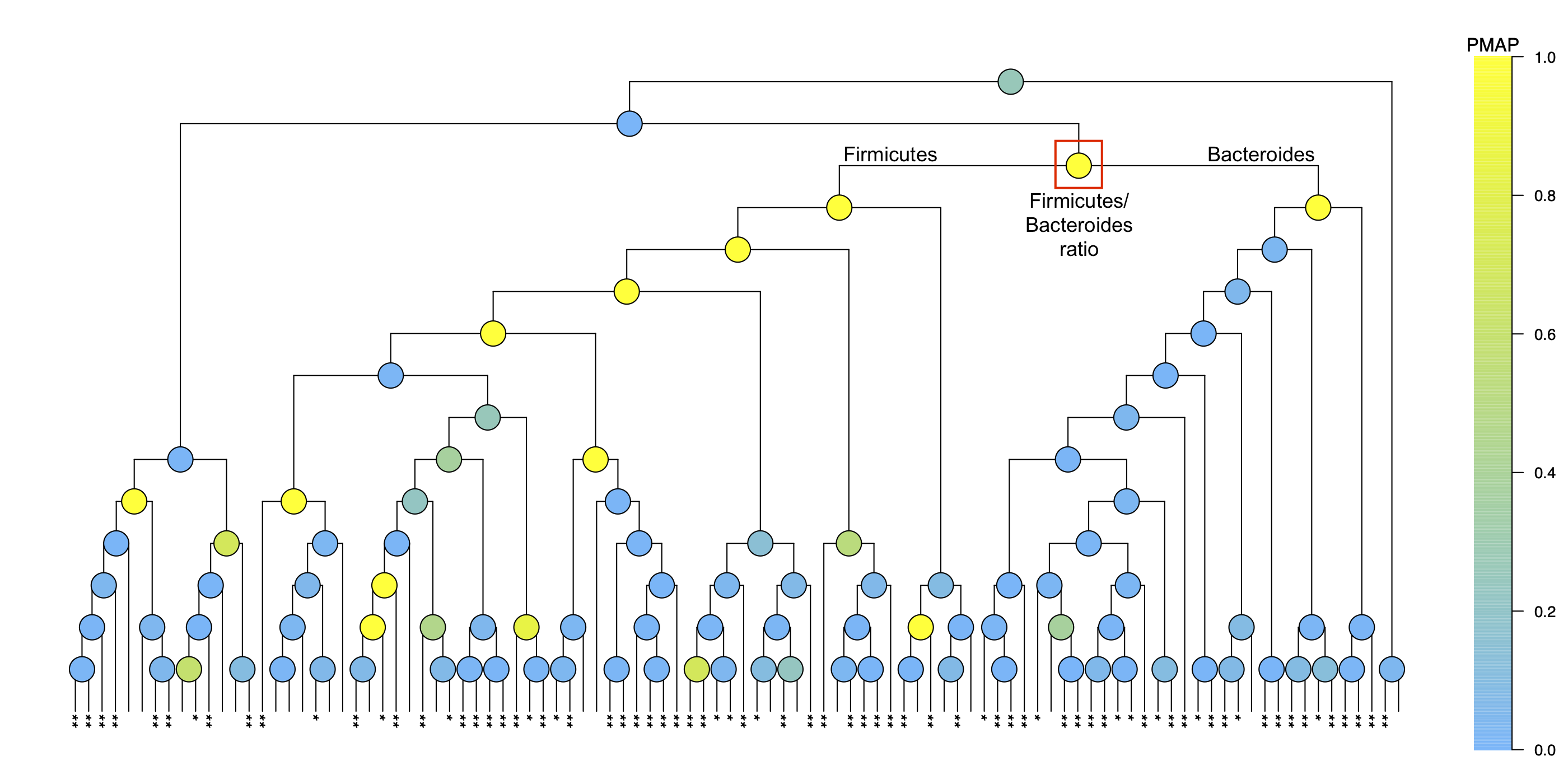}
    \caption{PMAPs of the introduction of solid food. The significant OTUs reported by MaAsLin2 are marked in the same way as Figure \ref{fig:pmap_qval_m005_seroconversion}. }
    \label{fig:solid_food}
\end{figure}

\begin{figure}[t]
    \centering
    \includegraphics[width=\textwidth]{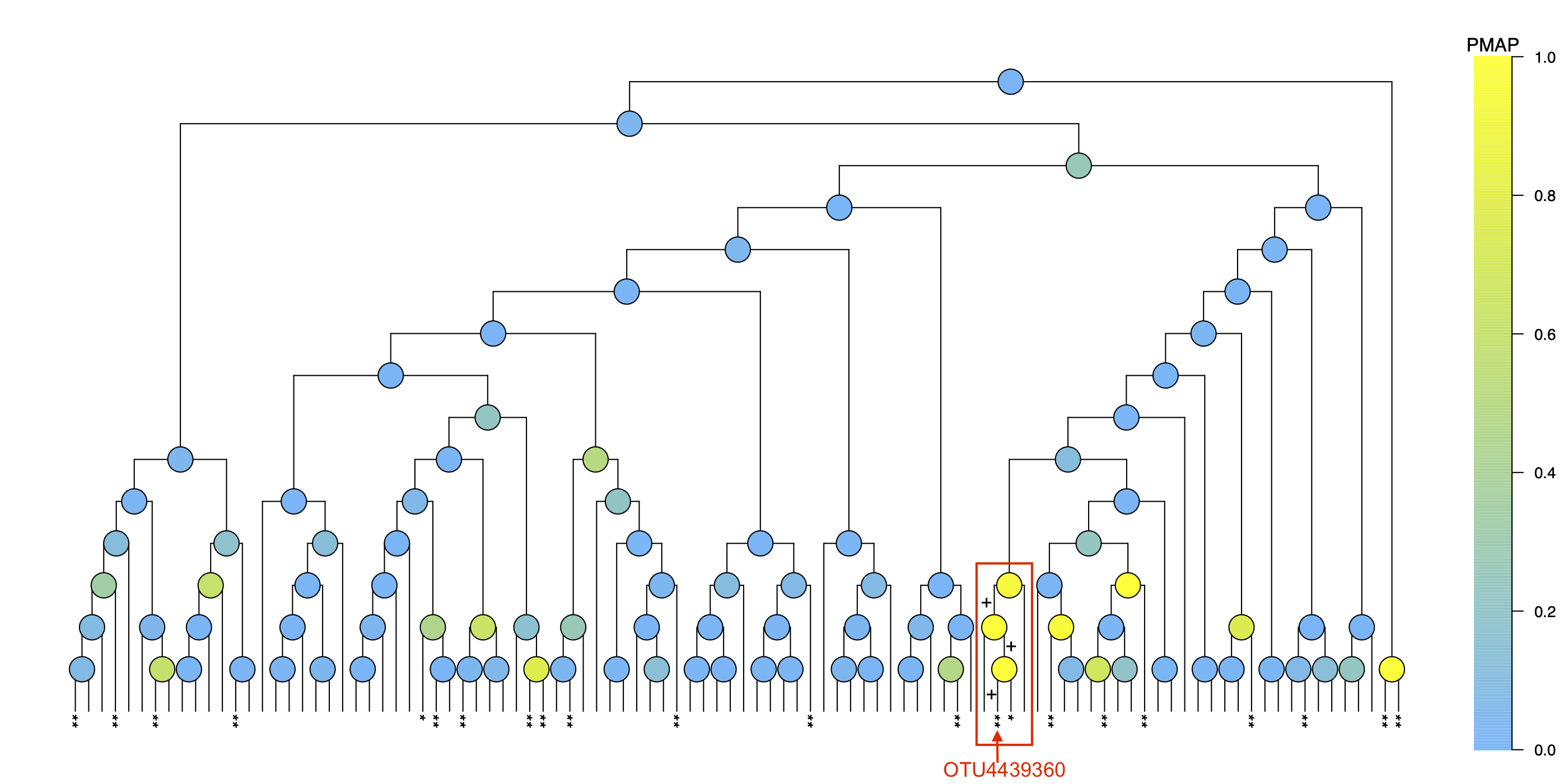}
    
    \caption{PMAPs of the introduction of soy products. The chain of three significant nodes and their descendant OTU 4439360 are marked in the figure. The posterior mean of $\alpha$ at these three nodes (from top to bottom) are 1.22, -3.12, and 3.19 respectively. The significant nodes reported by our LTN-based mixed-effects model and the significant OTUs reported by MaAsLin2 are marked in the same way as Figure \ref{fig:pmap_qval_m005_seroconversion}. }
    \label{fig:soy_prod}
\end{figure}

\begin{figure}
    \centering
    
    \includegraphics[width=\textwidth]{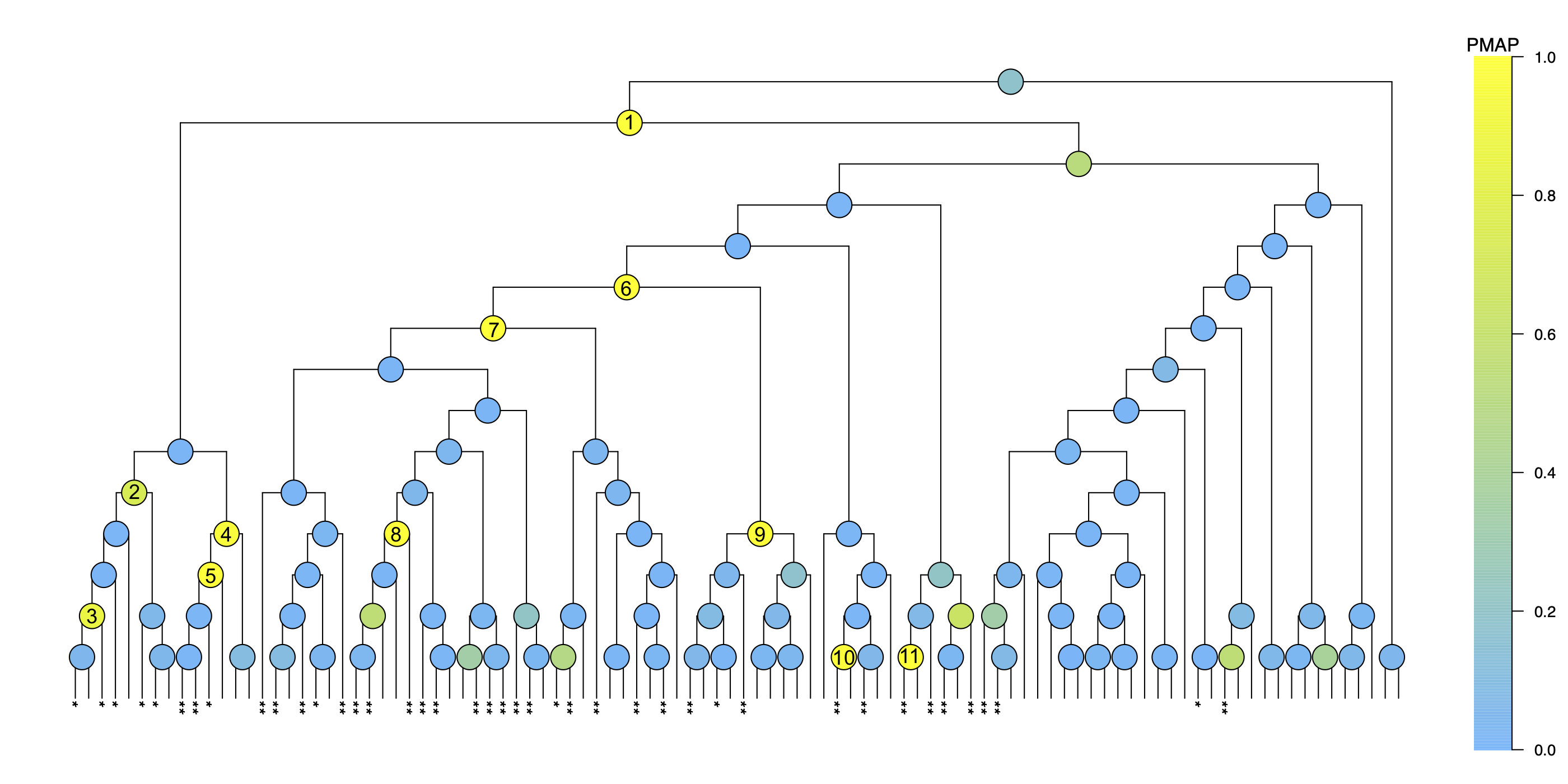}
    \caption{PMAPs of cessation of breastfeeding. The significant OTUs reported by MaAsLin2 are marked in the same way as Figure \ref{fig:pmap_qval_m005_seroconversion}. }
    \label{fig:bf}
\end{figure}

\begin{figure}[tbh]
    \centering
    \includegraphics[width=\linewidth]{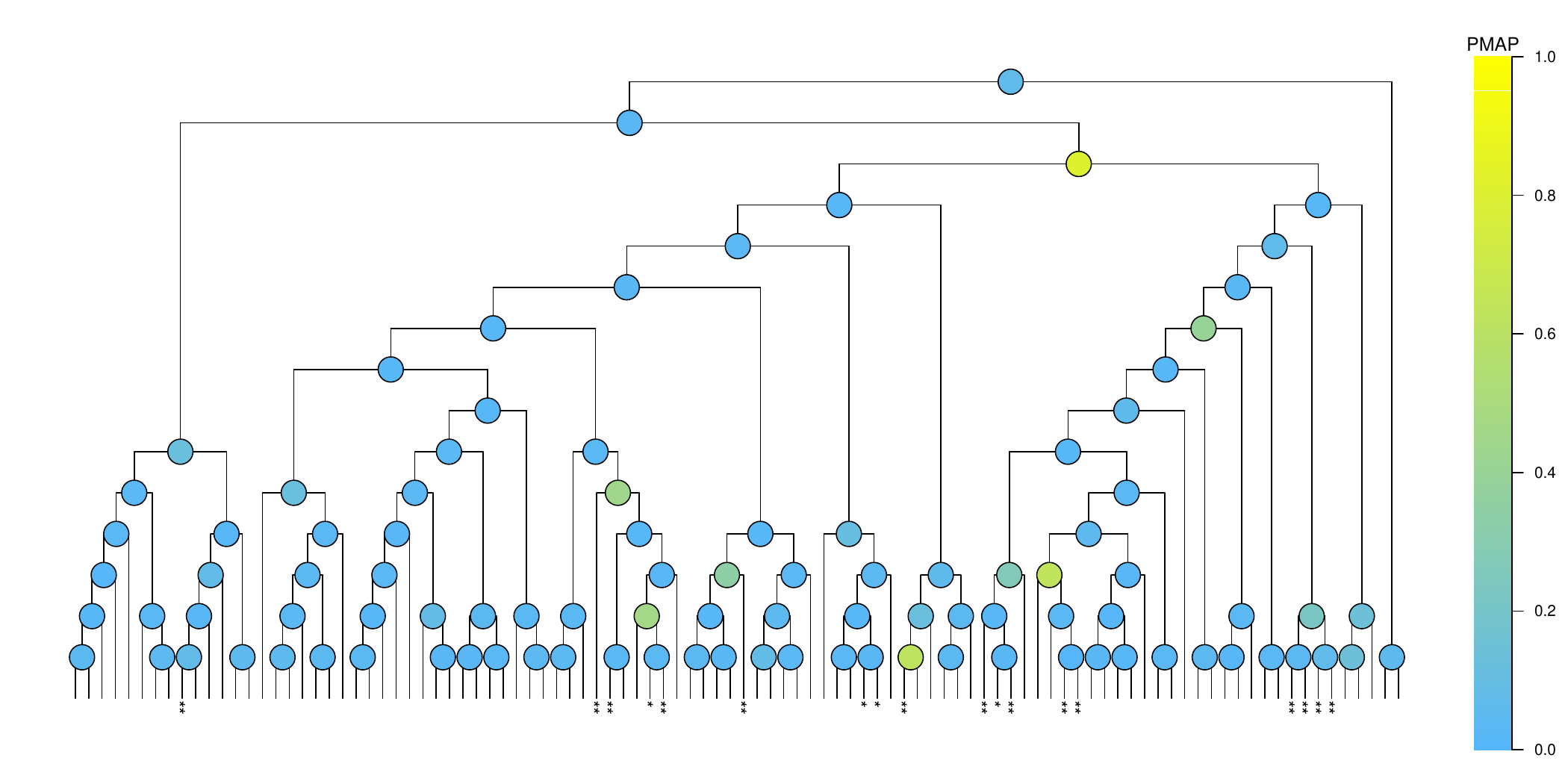}
    \caption{PMAPs for rye ($m=0.05$)}
    \label{fig:rye}
\end{figure}

\begin{figure}[tbh]
    \centering
    \includegraphics[width=\linewidth]{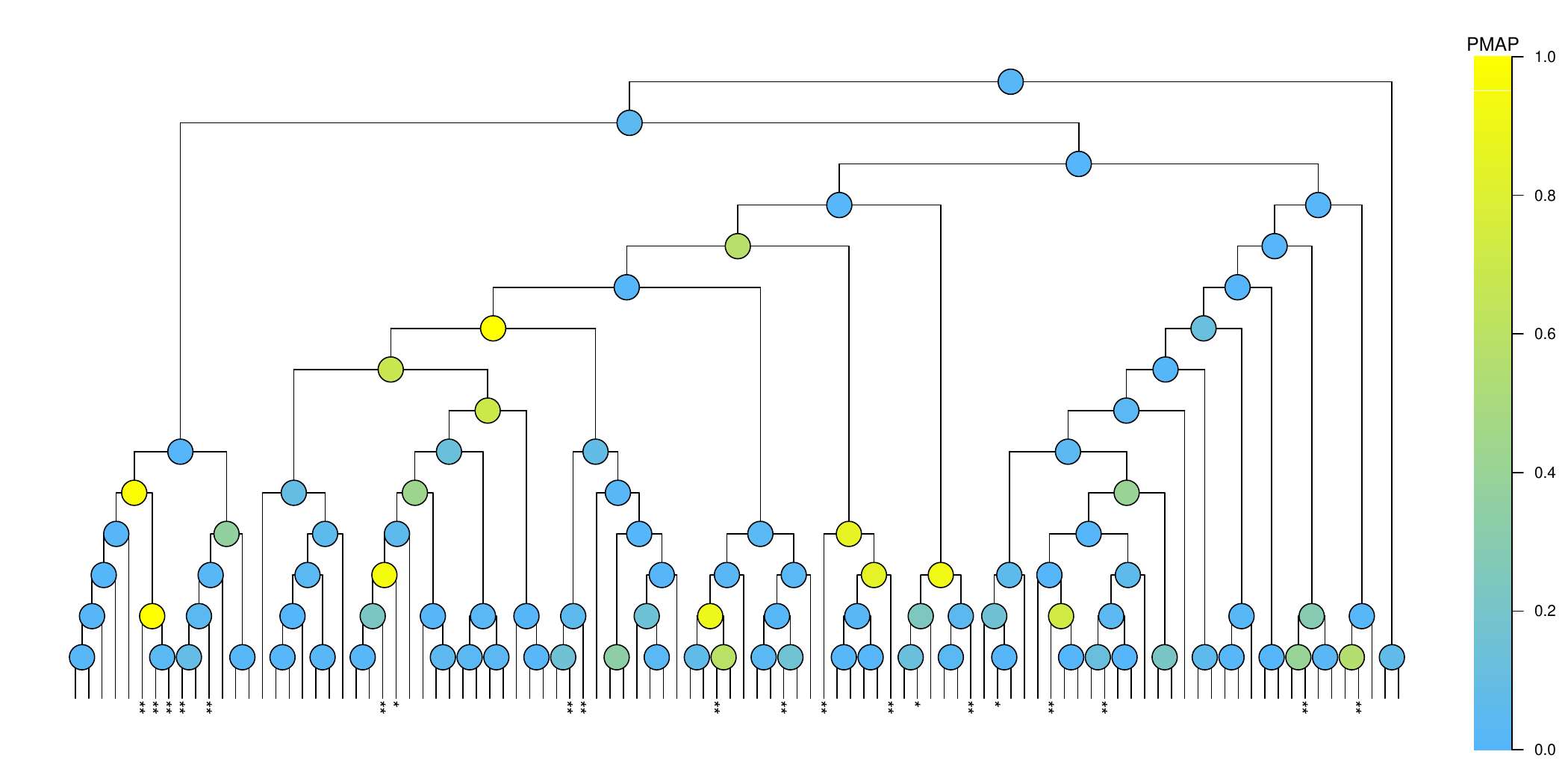}
    \caption{PMAPs for Buckwheat and Millet ($m=0.05$)}
    \label{fig:buckwheat}
\end{figure}

\begin{figure}[tbh]
    \centering
    \includegraphics[width=\linewidth]{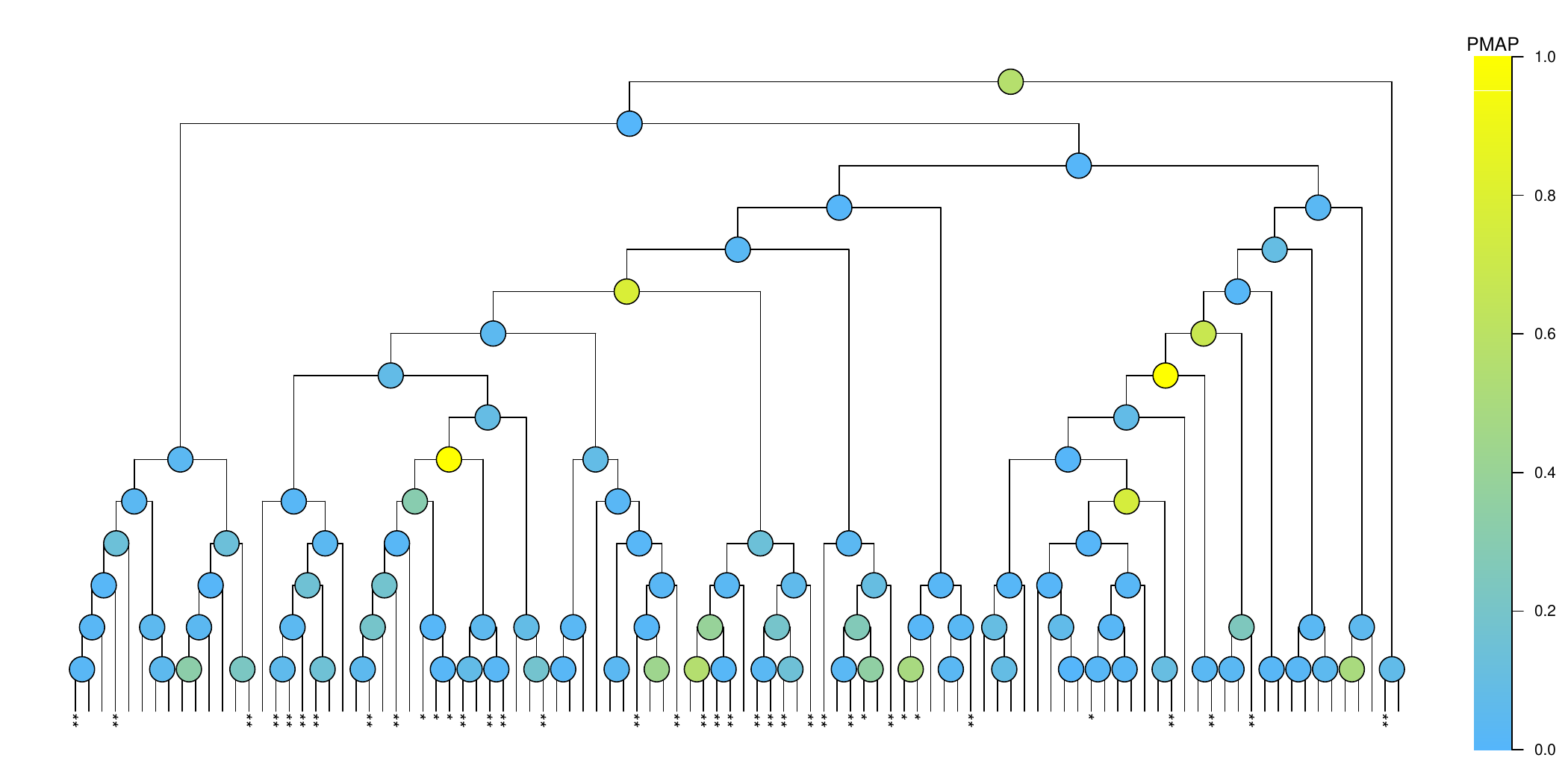}
    \caption{PMAPs for eggs ($m=0.05$)}
    \label{fig:eggs}
\end{figure}

\begin{figure}
    \centering
    \includegraphics[width=\linewidth]{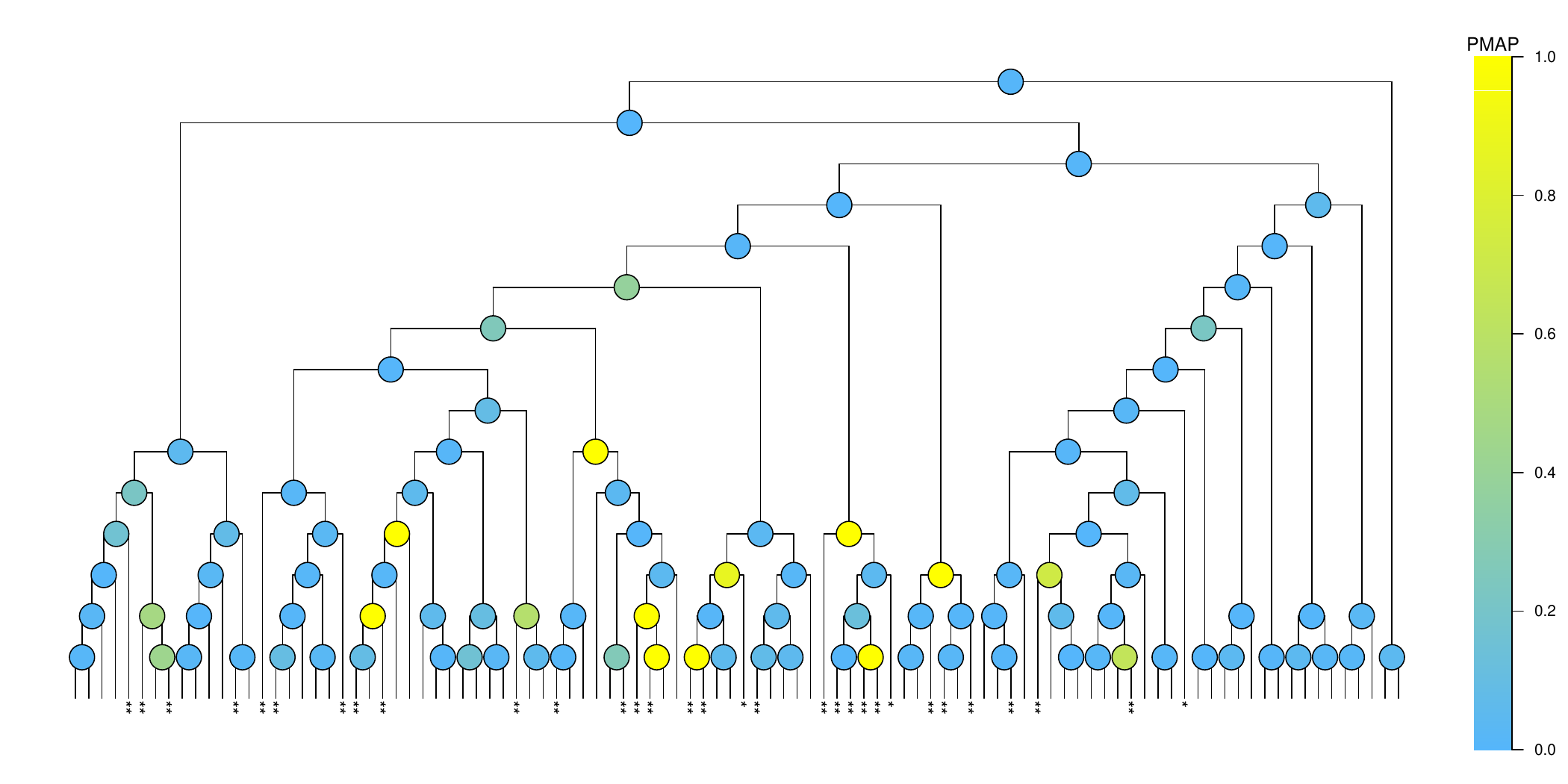}
    \caption{PMAPs for fish ($m=0.05$)}
    \label{fig:fish}
\end{figure}

\begin{figure}
    \centering
    \includegraphics[width=\linewidth]{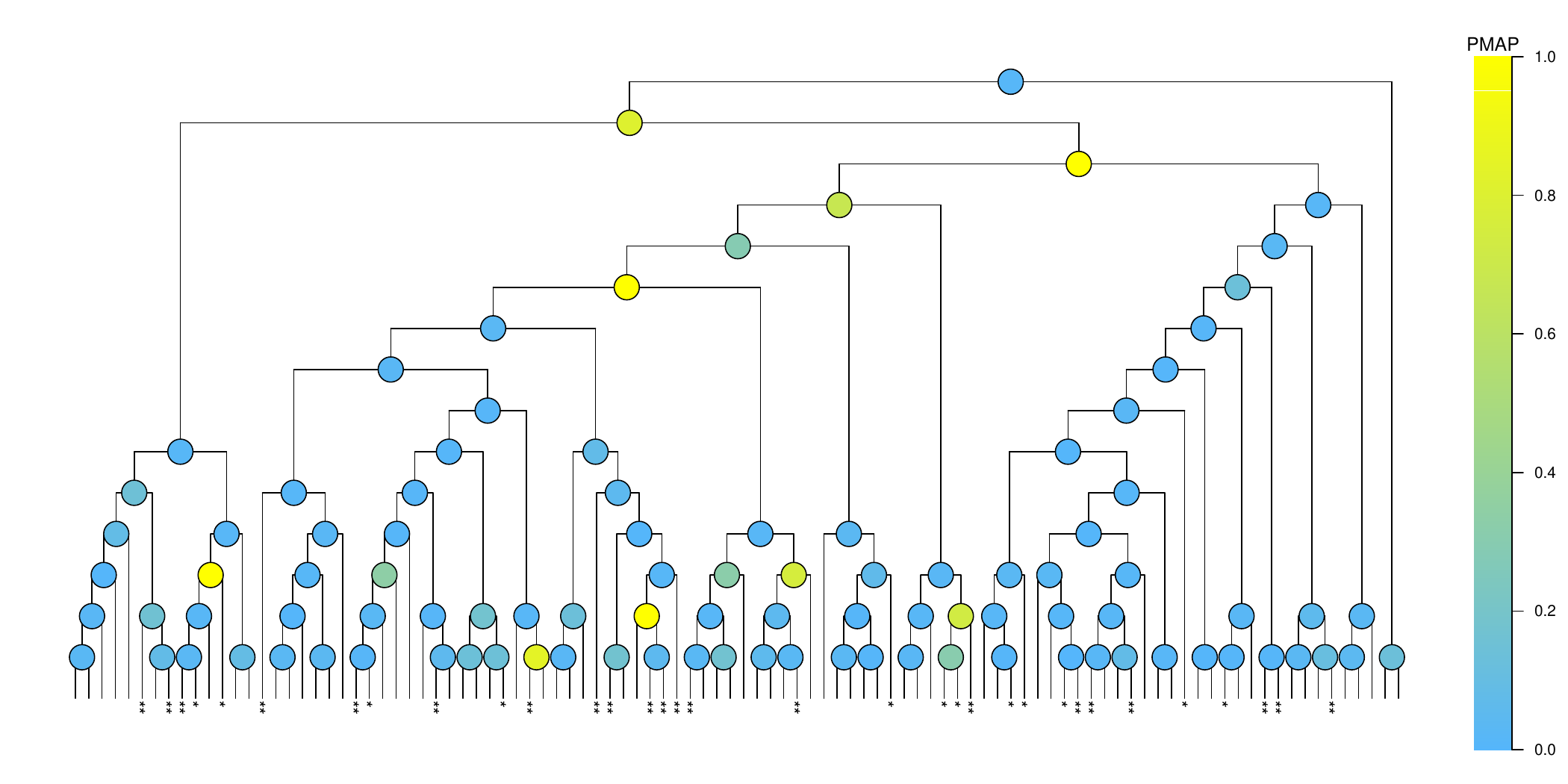}
    \caption{PMAPs for barley ($m=0.05$)}
    \label{fig:barley}
\end{figure}

The PMAPs (computed with $m=0.05$) are again visualized on the phylogenetic tree along with the significant OTUs reported under MaAsLin2 with BH-adjusted q-values for solid food (Figure~\ref{fig:solid_food}), soy products (Figure~\ref{fig:soy_prod}) and breastfeeding (Figure~\ref{fig:bf}). The PMAP plots for the other dietary variables are reported in Figures~\ref{fig:rye} - \ref{fig:barley}). 

Two distinct features of the tree-based parametrization are demonstrated in these figures. First, our model appears to lead to a sparser representation of the signals than the OTU-level analysis when cross-group differences are detected at multiple OTUs that are close to each other on the tree. For example, as shown in Figure \ref{fig:solid_food}, significant associations with solid food have been identified by our LTN-based mixed-effects model at relatively shallow nodes on the phylogenetic tree. In contrast, MaAsLin2 reported very large sets of significant descendant OTUs of these nodes. More interestingly, there is a node with high PMAP whose left and right child are Firmicutes and Bacteroidetes respectively (see Figure \ref{fig:solid_food}), indicating that introduction of solid food is associated with changes in the ``Firmicutes/Bacteroides ratio", which is widely used as an index of dysbiosis \citep{fb_ibd}. Second, ``chains'' of high PMAPs are observed in several comparisons, which can be useful in targeting the more precise taxa contributing to the cross-group difference. For example, in Figure \ref{fig:soy_prod} of the posterior marginal alternative probabilities of soy product, a chain of three ancestors of OTU 4439360 with the same direction of changes in the relative abundance implies higher relative abundance of OTU 4439360 after introducing soy product, which is consistent with data at older age (see Figure~\ref{fig:4439360}). 

Finally, we investigate how breastfeeding is associated with the gut microbiome development. In the comparison between samples collected before and after cessation of breastfeeding, the posterior marginal alternative probabilities are visualized in Figure \ref{fig:bf}, and for each significant node $A$, the taxa associated with the left and right child of this node, the common taxon they belong to, as well as the posterior mean $\hat{\alpha}(A)$, are summarized in Table \ref{table:bf}. The main discovery on breastfeeding in the original study by \cite{diabimmune} is an increase in Bifidobacterium and Lactobacillus species and reduction in Lachnospiraceae during breastfeeding. 
{ Table~\ref{table:bf} and Figure~\ref{fig:bf}  reveals some similar findings and more. 
The negative $\hat{\alpha}$ at node~7 indicates reduced Lachnospiraceae relative to Veillonellaceae in samples collected from infants during breastfeeding. Moreover, Bifidobacterium has higher relative abundance in the samples collected during breastfeeding period; indeed, the positive $\hat{\alpha}$ at nodes~4 and 5 (and their ancestors) indicates that such enrichment of Bifidobacterium can be attributed to longum and bifidum as well as some other unclassified species. Such result is consistent with previous findings on the high abundance of Bifidobacterium longum in breastmilk and enrichment of Bifidobacterium bifidum in breastfed infants \citep{breastmilk_prevalence,breastmilk_gut}.
}

\begin{table}[!ht]
    \centering
     \caption{Nodes with significant association with breastfeeding reported by our LTN-based mixed-effects model with posterior expected FDR $\leq$ 0.05. The node labels are the same as in Figure \ref{fig:bf}. }
    \begin{tabular}{p{1cm} p{3cm} p{4cm} p{3cm}p{2cm}}
    \hline
        node & taxon in common & taxa on left & taxa on right & posterior mean $\hat \alpha$ \\ \hline
        1 & Bacteria & Proteobacteria,\newline Actinobacteria & Firmicutes,\newline Bacteroidetes & 0.52 \\ 
        2 & Proteobacteria & Gammaproteobacteria & Betaproteobacteria & 0.94 \\ 
        3 & Enterobacteriaceae & OTU 782953,\newline OTU 668514 & OTU 2119418 & 1.28 \\
        4 & Actinobacteria & Actinobacteria & Coriobacteriia & 0.97 \\ 
        5 & Bifidobacterium &  longum, bifidum, \newline unclassified & adolescentis & 1.41 \\ 
        6 & Clostridiales & Lachnospiraceae, \newline Veillonellaceae & Ruminococcaceae & 1.11 \\
        7 & Clostridiales & Lachnospiraceae & Veillonellaceae & -1.61 \\ 
        8 & Lachnospiraceae & OTU 289734,\newline OTU 4483337,\newline OTU 2724175,\newline OTU 4448492 & OTU 4469576 & 1.35 \\
        9 & Ruminococcaceae & Oscillospira,\newline Faecalibacterium,\newline Ruminococcus & unclassified & -1.29 \\ 
        10 & Clostridiaceae & OTU 193672 & OTU 3576174 & -2.24 \\ 
        11 & Streptococcus & OTU 4442130 & OTU 4425214 & 1.50 \\ \hline
    \end{tabular}
   \label{table:bf}
\end{table}

\begin{figure}[tbh]
     \centering
     \begin{subfigure}[b]{0.45\textwidth}
         \centering
         \includegraphics[width=\textwidth]{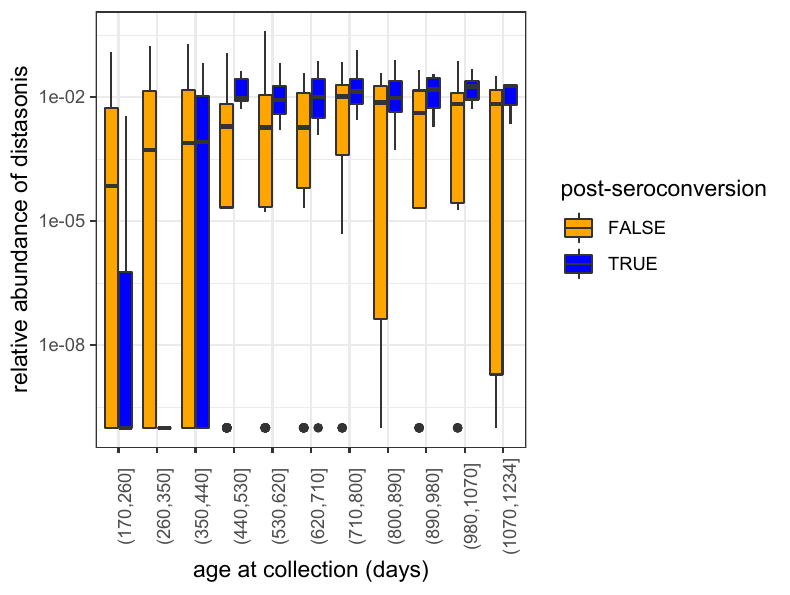}
         \caption{}
         \label{fig:distasonis}
     \end{subfigure}
     \hfill
     \begin{subfigure}[b]{0.45\textwidth}
         \centering
         \includegraphics[width=\textwidth]{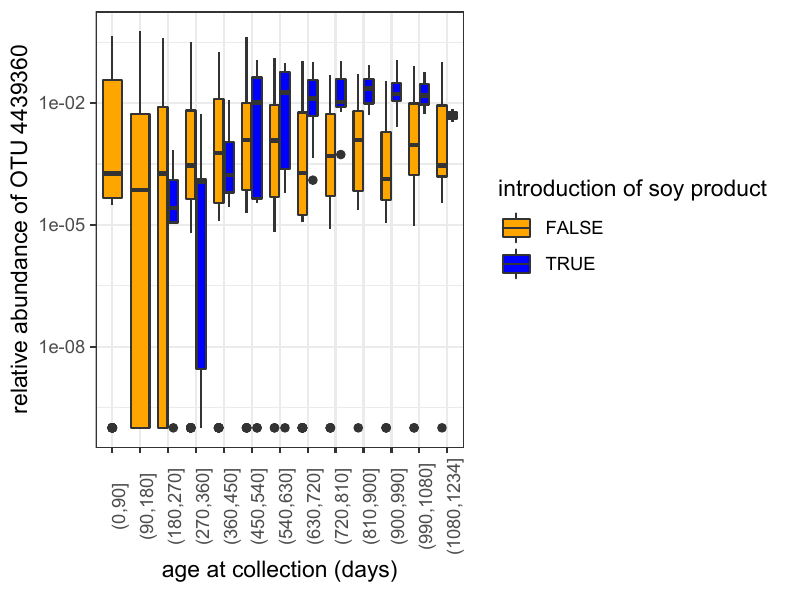}
         \caption{}
         \label{fig:4439360}
     \end{subfigure}
        \caption{(a) Relative abundance of distasonis. Data collected before the first event of seroconversion is omitted. (b) Relative abundance of OTU 4439360.}
\end{figure}

\end{document}